\documentclass{jfm}
\usepackage{amsmath,amsfonts,amssymb,amsgen,amsbsy,upmath,bm,graphicx,natbib}

\newcommand{\ud}{\mathrm{d}}
\newcommand{\ue}{\mathrm{e}}
\newcommand{\ui}{\mathrm{i}}

\newcount\ndots
 \def\drawline#1#2{\raise 2pt\vbox{\hrule width #1pt height #2pt}}
 \def\spacce#1{\hskip #1pt}
 \def\solid{\drawline{16}{.1}\nobreak}
 \def\bdash{\hbox{\drawline{2}{.1}\spacce{2}}}
 \def\dashed{\bdash\bdash\bdash\bdash\bdash\nobreak}

\title[Damped~oscillations of~elastic~quasi-circular~membrane]{On~the~damped~oscillations of~an~elastic~quasi-circular~membrane in~a~two-dimensional~incompressible~fluid}
\author[Marco Martins Afonso, Simon Mendez \& Franck Nicoud]{Marco Martins Afonso, Simon Mendez and Franck Nicoud}
\affiliation{Institut~de~Math\'ematiques~et~de~Mod\'elisation~de~Montpellier, CNRS UMR 5149, Universit\'e~Montpellier~2, c.c.051, 34095~Montpellier~cedex~5, France}
\date{31/5/2013}

\begin{document}
\maketitle

\begin{abstract}
 We propose a procedure --- partly analytical and partly numerical ---
 to find the frequency and the damping rate of the small-amplitude oscillations
 of a massless elastic capsule immersed in a two-dimensional viscous incompressible fluid.
 The unsteady Stokes equations for the stream function
 are decomposed onto normal modes for the angular and temporal variables,
 leading to a fourth-order linear ordinary differential equation in the radial variable.
 The forcing terms are dictated by the properties of the membrane,
 and result into jump conditions at the interface between the internal and external media.
 The equation can be solved numerically,
 and an excellent agreement is found with a fully-computational approach we developed in parallel.
 Comparisons are also shown with the results available in the scientific literature for drops,
 and a model based on the concept of embarked fluid is presented,
 which allows for a good representation of the results and a consistent interpretation of the underlying physics.
\end{abstract}

\section{Introduction} \label{intr}

 During the last two decades, numerical simulations of deformable particles in flows have received increasing attention.
 Several numerical methods have been developed to predict the dynamics of particles composed by fluid enclosed in a membrane, or more
 generally in a flexible structure.
 Different particles are described by this model, such as biological cells (typically red blood cells), lipid vesicles or elastic capsules.
 Such particles are small (their diametre is of the order of a few microns), and they often evolve in creeping flows,
 well described by the Stokes equations. Taking advantage of the linearity of the Stokes equations,
 boundary integral methods (BIM) \citep{Pozrikidis:1992,Pozrikidis:2010} are efficient techniques to solve the dynamics of fluidic particles
 in an external flow. These methods have been successfully applied to red blood cells \citep{Pozrikidis:1995,Zhao:2010,Zhao:2011a},
 capsules \citep{Lac:2007,Walter:2011,Hu:2012} and vesicles \citep{Ghigliotti:2010a,Veerapaneni:2011,Biben:2011,Boedec:2012}.
 An appealing feature of BIM is their precision, as they do not need the discretisation of the fluid domain \citep{Pozrikidis:1992}.

 However, as recently stressed by \cite{Salac:2012}, vesicles (and other similar cells) sometimes evolve in flows where inertia cannot
 be neglected. As far as red blood cells are concerned, the particle Reynolds number is generally small under normal physiological conditions.
 However, non-physiological flows are prime examples where cells are submitted to extreme stress conditions. In ventricular assist devices,
 the Reynolds number can reach $10^4\div10^5$ \citep{Fraser:2011}, potentially resulting in blood damage. In such conditions, the flow cannot
 be described by the Stokes equations, and BIM cannot be used.

 Numerical methods able to predict flow-induced deformations of fluid particles
 enclosed by a membrane in regimes where the Reynolds number is not zero have thus been developed
 in recent years \citep{Peskin:2002,Lee:2003,Le:2006,Cottet:2006,Kim:2010,Salac:2011}. Such methods are usually developed
 and tested in two dimensions before their adaptation to three dimensions.
 Quality assessment of these methods is of course essential. However, validation itself is an issue:
 while an experiment may mimic a two-dimensional flow, flowing particles are inherently 3D.
 Likewise, owing to the moderate computational cost
 of BIM, reference numerical results generated by BIM are 3-D for the most part.
 The comparison of 2-D results and 3-D measurements or simulation results can only be qualitative, and does not constitute a true validation.
 However, some 2-D simulations using BIM are available in the literature
 \citep{Breyiannis:2000,Veerapaneni:2009a,Ghigliotti:2010a,Woolfenden:2011}
 and can be used for validation, in the Stokes regime. Validation in regimes where the flow cannot be described by the Stokes equation
 is more problematic. In such cases, numerical publications are generally limited to simulation feasibility and qualitative assessments,
 or comparisons only with similar methods:
 most of the methods developed to compute capsules or vesicles dynamics
 in flows with non-zero Reynolds number rely on diffuse interfaces, where the effect of the membrane is smoothed over several grid cells.
 Extensive validation of such methods is thus expected, and analytical results --- necessary to properly discuss the pros and cons of new
 numerical methods --- are missing.

 The present paper represents an analytical attack to a problem already faced several times numerically in literature,
 and is a step towards more extensive validation of 2-D numerical methods for simulations of fluid particle dynamics under flow.
 A normal-mode analysis is performed for the case of a two-dimensional capsule, enclosed by an elastic stretched membrane. A perturbation of the
 membrane equilibrium shape results in damped oscillations and relaxation to a circular state. Such a case has been used
 to illustrate the performances of 2-D numerical methods \citep{Tu:1992,Lee:2003,Cottet:2006,Song:2008,Tan:2008}, albeit without validating against
 analytical results.

 Similar studies have been led for other deformable objects.
 The problem of oscillations of drops or vesicles has received substantial attention over the years \citep{Lamb:1932,Prosperetti:1980a,Lu:1991,Rochal:2005}.
 However, most of the approaches treat three-dimensional configurations, relevant of course to practical applications.
 The present paper provides semi-analytical results for the oscillating relaxation of two-dimensional capsules without external flow,
 and proposes a discussion on the similarities with the behaviour of droplets.
 We will also introduce a simple phenomenological model to explain the origin of the oscillations of a massless membrane immersed in an inertial fluid.

 The paper is organised as follows.
 In section \ref{prob} we sketch the problem under consideration.
 Section \ref{equa} shows the guidelines of the equations into play.
 In section \ref{resu} the main results are reported.
 Section \ref{disc} presents a discussion on similar systems and on physical models.
 Conclusions and perspectives follow in section \ref{conc}.
 Appendix \ref{appe} is devoted to detailing the analytical procedure.

\section{Problem} \label{prob}

\begin{figure}
 \centering
 \includegraphics[width=6.5cm]{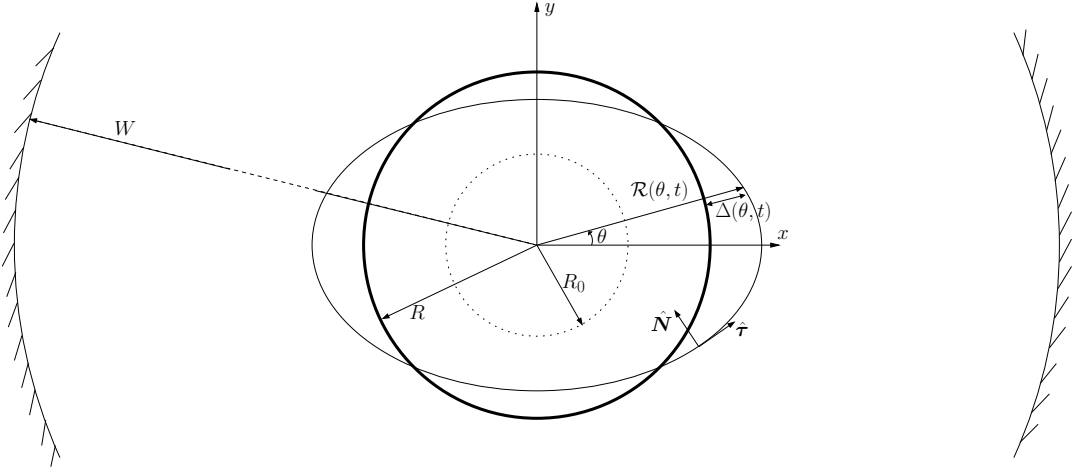}\vskip0.5cm
 \caption{Sketch of the problem and of the notations for membrane, fluid and wall.
  The inflated membrane with radius $R$ at rest oscillates within a circular fluid domain of radius $W$.
  The dotted circle represents the zero-stress configuration of the membrane of radius $R_0$.}
 \label{sche}
\end{figure}

 Let us consider the motion of an elastic one-dimensional closed membrane, slightly deformed from its circular rest shape,
 immersed in --- and thus also enclosing a portion of --- a two-dimensional incompressible viscous fluid (figure \ref{sche}).
 We study the time evolution of the membrane shape, which consists in time-decaying oscillations around the reference circle
 enclosing the same area as the initial form (which, for sake of visual simplification, can be thought of as
 similar to an ellipse). As we focus on the case of small-amplitude deformations,
 we can describe the problem in polar coordinates $\bm{r}=(r,\theta)$ via the membrane location $\mathcal{R}(\theta,t)=R+\Delta(\theta,t)$,
 where $R$ is the radius of the aforementioned circle and $\Delta(\theta,t)/R\ll1\ \forall\;\theta,t$.
 This description of course neglects the membrane thickness and excludes those situations where the membrane folds, so that two or more
 membrane locations would be present at the same angle. We also introduce a long-wave hypothesis, in order to neglect
 the cases in which the instantaneous oscillations would be so spatially dense as to result in a quasi-radial orientation of the interface.
 On the contrary, we take into account the possibility of pre-inflating the membrane --- from an initial radius $R_0<R$
 to the reference one $R$ --- before deforming it. Consequently, the local instantaneous curvature is defined as \citep{Gray:1997}
 \begin{equation} \label{curv}
  \Gamma(\theta,t)=\frac{\mathcal{R}^2+2\mathcal{R}'^2-\mathcal{R}\mathcal{R}''}{(\mathcal{R}^2+\mathcal{R}'^2)^{3/2}}=\frac{(R+\Delta)^2+2\Delta'^2-(R+\Delta)\Delta''}{[(R+\Delta)^2+\Delta'^2]^{3/2}}\;,
 \end{equation}
 with the prime denoting a derivative with respect to $\theta$.

 In the spirit of a perturbative approach, the continuity and forced Navier--Stokes equations for the fluid velocity
 can be linearised around the reference state of fluid at rest,
 and the resulting equations for the perturbation $\bm{u}(r,\theta,t)=u_r\hat{\bm{e}}_r+u_{\theta}\hat{\bm{e}}_{\theta}$ read:
 \begin{equation} \label{cont}
  \frac{1}{r}\frac{\partial(ru_r)}{\partial r}+\frac{1}{r}\frac{\partial u_{\theta}}{\partial\theta}=0\;,
 \end{equation}
 \begin{equation} \label{nsrad}
  \rho\frac{\partial u_r}{\partial t}=-\frac{\partial p}{\partial r}+\nu\left(\frac{\partial^2 u_r}{\partial r^2}+\frac{1}{r}\frac{\partial u_r}{\partial r}+\frac{1}{r^2}\frac{\partial^2u_r}{\partial\theta^2}-\frac{u_r}{r^2}-\frac{2}{r^2}\frac{\partial u_{\theta}}{\partial\theta}\right)+\delta(r-\mathcal{R})f_r\;,
 \end{equation}
 \begin{equation} \label{nsazi}
  \rho\frac{\partial u_{\theta}}{\partial t}=-\frac{1}{r}\frac{\partial p}{\partial\theta}+\nu\left(\frac{\partial^2 u_{\theta}}{\partial r^2}+\frac{1}{r}\frac{\partial u_{\theta}}{\partial r}+\frac{1}{r^2}\frac{\partial^2u_{\theta}}{\partial\theta^2}-\frac{u_{\theta}}{r^2}+\frac{2}{r^2}\frac{\partial u_r}{\partial\theta}\right)+\delta(r-\mathcal{R})f_{\theta}\;,
 \end{equation}
 where $\rho$ and $\nu$ are the (constant) mass density and kinematic viscosity, respectively, $p$ is the fluid pressure,
 and the nonlinear inertial terms (quadratic coupling of $\bm{u}$ with itself) have been consistently dropped.\\
 The linear force per unit volume, $\bm{f}(\theta,t)=f_r\hat{\bm{e}}_r+f_{\theta}\hat{\bm{e}}_{\theta}$, exerted by the membrane on the fluid,
 can easily be expressed in the reference frame defined by the unit vectors
 locally counterclockwise-tangential and inward-normal to the interface,
 $\hat{\bm{\tau}}$ and $\hat{\bm{N}}$ respectively (see figure \ref{sche}).
 In terms of the local instantaneous elongation $\chi(\theta,t)$ of the membrane,
 this force reads \citep{Breyiannis:2000}
 \begin{equation} \label{force}
  \bm{f}=E\left(\frac{\ud\chi}{\ud\tau}\hat{\bm{\tau}}+\Gamma\chi\hat{\bm{N}}\right)\;,
 \end{equation}
 where $\ud/\ud\tau$ is the curvilinear derivative following $\hat{\bm{\tau}}$,
 and $E$ is an elastic modulus representing the membrane thickness
 times its Young modulus and is considered as constant in the limit of vanishing thickness.\\
 The membrane elongation is defined as
 \begin{equation} \label{elref}
  \chi(\theta,t)=\frac{L(\theta,t)-L_0}{L_0}\;,
 \end{equation}
 $L$ being the length of an infinitesimal membrane arc and $L_0$ its deflated value,
 which will actually disappear from the final expression.
 As shown in appendix \ref{appe}, such a definition leads to an expression dependent not only on the geometric configuration,
 but also on the dynamical state, through the appearance of the fluid velocity.\\
 Moreover, the Dirac deltas in (\ref{nsrad}--\ref{nsazi}) indicate that this force only acts at the local and instantaneous
 membrane location, $r=\mathcal{R}(\theta,t)$. However, in the limit of small-amplitude deformations,
 this differs only slightly from the circle
 $r=R$, and at leading order all the relevant expressions (and consequently (\ref{force}))
 can be simplified by expanding them around the reference state. The details of the calculation can be found in the appendix.\\
 Expression (\ref{force}) incorporates both the normal force related to fluid pressure differences,
 and the tangential stress due to variations in the membrane local elongation.
 It is thus obvious that this will result in jump conditions, which in the same spirit
 should be imposed on the \emph{fixed and known} reference circular state. They must be supplemented by the continuity of the velocity field,
 \begin{equation} \label{jump}
  u_r^-=u_r^+\qquad\textrm{and}\qquad u_{\theta}^-=u_{\theta}^+\;,
 \end{equation}
 where $\cdot^{\pm}=\lim_{r\to R_{\pm}}\cdot$ and all equalities hold for any angle and time.
 The forthcoming partial differential equations will then hold separately
 in both the internal and the external fluid regions, and will have to be matched according with (\ref{jump}).\\
 Finally, boundary conditions must be imposed: both at the origin due to the polar-coordinate geometric constraint,
 and at the external wall (which may even be placed at infinity, but which in what follows will be considered
 as a concentric circle of radius $W>R$, for the sake of simplicity) because of the no-slip constraint:
 \begin{equation} \label{bc}
  u_r|_{r=0}=0=\left.\frac{\partial u_r}{\partial r}\right|_{r=0}\;,\qquad u_r|_{r=W}=0=u_{\theta}|_{r=W}\;.
 \end{equation}

 This completes the basic picture of the physical system under investigation.
 It is however instructive to recast the problem in nondimensional form.
 With this aim, let us introduce \citep{Prosperetti:1980a} the membrane characteristic time
 \begin{equation} \label{T}
  T=\sqrt{\frac{\rho}{E}\frac{R^3R_0}{R-R_0}}\;,
 \end{equation}
 and the so-called reduced viscosity, corresponding to the ratio between $T$ itself and the viscous time scale $R^2/\nu$:
 \begin{equation} \label{Sm}
  \nu_*=\frac{\nu T}{R^2}=\sqrt{\frac{\nu\mu R_0}{ER(R-R_0)}}
 \end{equation}
 (here $\mu=\rho\nu$ is the dynamic viscosity).
 A standard application of the well-known Vaschy--Buckingham theorem $\pi$ tells us that
 the problem can be formulated in terms of two nondimensional quantities, which can be chosen as
 $\nu_*$ and the ratio $R_*= R_0/R$, as we will show at the end of section \ref{equa}.

\section{Equations} \label{equa}

 Let us obviously start our analysis from the linearised forced Navier--Stokes equations.
 By multiplying (\ref{nsazi}) by $r$, differentiating it with respect to $r$, and subtracting it from the equation
 obtained by differentiating (\ref{nsrad}) with respect to $\theta$, we get rid of pressure in the equation:
 \begin{eqnarray} \label{mixed}
  \frac{\partial^2u_r}{\partial\theta\partial t}-\frac{\partial}{\partial r}\left(r\frac{\partial u_{\theta}}{\partial t}\right)&=&\nu\left\{\frac{\partial}{\partial\theta}\left[\frac{\partial^2 u_r}{\partial r^2}+\frac{1}{r}\frac{\partial u_r}{\partial r}+\frac{1}{r^2}\frac{\partial^2u_r}{\partial\theta^2}-\frac{u_r}{r^2}-\frac{2}{r^2}\frac{\partial u_{\theta}}{\partial\theta}\right]\right.\nonumber\\
  &&\quad\left.-\frac{\partial}{\partial r}\left[r\frac{\partial^2 u_{\theta}}{\partial r^2}+\frac{\partial u_{\theta}}{\partial r}+\frac{1}{r}\frac{\partial^2u_{\theta}}{\partial\theta^2}-\frac{u_{\theta}}{r}+\frac{2}{r}\frac{\partial u_r}{\partial\theta}\right]\right\}\nonumber\\
  &&+\frac{E}{\rho R_0}\left\{\left[\Delta'\left(1-\frac{R_0}{R}\right)\left(1-\frac{r}{R}\right)-\frac{rR_0}{R}\chi'\right]\dot{\delta}(r-R)\right.\nonumber\\
  &&\qquad+\left.\left[\frac{\Delta'''}{R}\left(1-\frac{R_0}{R}\right)-2\frac{R_0}{R}\chi'\right]\delta(r-R)\right\}\;.
 \end{eqnarray}
 The superscript dot here denotes a derivative with respect to the argument itself of the delta.
 (Notice that (\ref{mixed}) represents the equation for the vorticity field, multiplied by $-r$).\\
 The continuity equation (\ref{cont}) is automatically satisfied by introducing the stream function $\psi(r,\theta,t)$ such that
 \begin{equation} \label{stream}
  u_r=\frac{1}{r}\frac{\partial\psi}{\partial\theta}\;,\qquad u_{\theta}=-\frac{\partial\psi}{\partial r}\;.
 \end{equation}

 We now perform a normal-mode analysis, i.e.\ we project onto a basis made up of complex exponentials in the angle and in time, in the form
 \begin{equation} \label{nom}
  \psi(r,\theta,t)=\sum_{n=-\infty}^{+\infty}\phi_n(r)\ue^{\ui(n\theta-\omega_nt)}\;,
 \end{equation}
 coupled with an analogous expansion for the two relevant quantities pertaining to the membrane,
 viz.\ its deformation $\Delta$ and its elongation $\chi$.
 The real and imaginary parts of $\omega_n$ correspond to the angular frequency of the membrane oscillations
 and to the (opposite of the) damping rate due to viscous dissipative effects, respectively, for the mode $n$.
 Finding this complex parametre --- common to both the membrane and the fluid --- is our main objective,
 which also means that we renounce to impose any initial condition for the fluid motion, in the spirit of an eigenmode decomposition.
 Both for this angular frequency and for the amplitudes $\phi_n(r)$ we drop the subscript $n$ whenever unambiguous. 

 Postponing the details to the appendix, here we only mention that, after the appropriate substitutions,
 equation (\ref{mixed}) can be recast in closed form by expressing the link between the membrane-related quantities
 $\Delta$, $\chi$ and the fluid stream function $\psi$ (or $\phi$) itself.
 As a result, in each of the two domains $r<R$ and $r>R$, we have
 \begin{eqnarray} \label{each}
  \phi^{(4)}+\frac{2}{r}\phi^{(3)}+\left(-\frac{2n^2+1}{r^2}+\frac{\ui\omega}{\nu}\right)\phi^{(2)}+\left(\frac{2n^2+1}{r^3}+\frac{\ui\omega}{\nu r}\right)\phi^{(1)}&\nonumber\\
  +\left(\frac{n^4-4n^2}{r^4}-\frac{\ui\omega n^2}{\nu r^2}\right)\phi&=&0\;,
 \end{eqnarray}
 with $^{(l)}$ denoting the $l$-th derivative with respect to $r$.\\
 This must be supplemented with an appropriate rewriting of the constraints (\ref{jump}) and (\ref{bc}).
 The boundary conditions at the origin and at the (circular) wall imply the vanishing
 of both the function and of its first derivative at both ends, and read:
 \begin{equation} \label{bcf}
  \phi|_{r=0}=0=\phi|_{r=W}\;,\qquad\phi^{(1)}|_{r=0}=0=\phi^{(1)}|_{r=W}\;.
 \end{equation}
 The continuity of the fluid velocity at the membrane implies the same for the function and its first derivative:
 \begin{equation} \label{mem1}
  \phi^-=\phi^+\;,\qquad\phi^{(1)-}=\phi^{(1)+}\;.
 \end{equation}
 Any constraint about pressure is no more relevant in this formulation.
 However, the effect of the Dirac delta's from (\ref{nsrad}--\ref{nsazi}) is to provide additional jump conditions
 on the second and third derivative of $\phi$ at the interface,
 which complete (\ref{bcf}-\ref{mem1}) and can be derived by means of appropriate successive integrations
 (see appendix \ref{appe} for more details):
 \begin{equation} \label{mem2}
  \phi^{(2)-}-\phi^{(2)+}=\frac{\ui n^2E}{\mu R_0R^2\omega}\left[\phi(R)-R\phi^{(1)}(R)\right]\;,
 \end{equation}
 \begin{equation} \label{mem3}
  \phi^{(3)-}-\phi^{(3)+}=\frac{\ui n^2(n^2-1)E}{\mu R_0R^3\omega}\left(1-\frac{R_0}{R}\right)\phi(R)\;.
 \end{equation}

 Finally, let us first reformulate the problem in nondimensional fashion,
 as anticipated at the end of section \ref{prob}. In other words, by using the nondimensionalised
 radial coordinate $r\mapsto r_*= r/R$ and angular frequency $\omega\mapsto\omega_*=\omega T$,
 and by exploiting (\ref{T}--\ref{Sm}), equation (\ref{each}) and constraints (\ref{mem2}--\ref{mem3}) can be recast as:
 \begin{eqnarray} \label{nondi}
  \phi^{(4)}(r_*)+\frac{2}{r_*}\phi^{(3)}(r_*)+\left(-\frac{2n^2+1}{r_*^2}+\frac{\ui\omega_*}{\nu_*}\right)\phi^{(2)}(r_*)&&\nonumber\\
  +\left(\frac{2n^2+1}{r_*^3}+\frac{\ui\omega_*}{\nu_*r_*}\right)\phi^{(1)}(r_*)+\left(\frac{n^4-4n^2}{r_*^4}-\frac{\ui\omega_*n^2}{\nu_*r_*^2}\right)\phi(r_*)&=&0\;,
 \end{eqnarray}
 \begin{equation} \label{non2}
  \phi^{(2)-}-\phi^{(2)+}=\frac{\ui n^2}{\omega_*\nu_*(1-R_*)}\left[\phi(1)-\phi^{(1)}(1)\right]\;,
 \end{equation}
 \begin{equation} \label{non3}
  \phi^{(3)-}-\phi^{(3)+}=\frac{\ui n^2(n^2-1)}{\omega_*\nu_*}\phi(1)\;.
 \end{equation}
 Notice that, due to the homogeneity of (\ref{nondi}) \& (\ref{non2}--\ref{non3}) in $\phi$, it is not necessary
 to nondimensionalise this latter quantity, of dimension length squared over time.
 This means that, in units of $T^{-1}$, the oscillation frequency and the damping rate can be expressed as
 \begin{equation} \label{omegas}
  \Re(\omega_*)=h(n,\nu_*,R_*)\;,\qquad\Im(\omega_*)=g(n,\nu_*,R_*)
 \end{equation}
 by means of two unknown functions $h$ and $g$, whose behaviour will be investigated numerically in section \ref{resu}.

\section{Results} \label{resu}

 In this section we show the analytical/numerical results for the real and imaginary parts of the nondimensional $\omega_*$.
 A numerical procedure is indeed necessary to obtain the nondimensional functions $h$ and $g$ from (\ref{omegas}).
 A standard finite-difference scheme is implemented, consisting in discretising the $r_*\in[0,W/R]$ domain with $M$ points,
 in transforming (\ref{nondi}) into a set of $M$ coupled linear equations accordingly
 (with conditions (\ref{bcf}--\ref{mem1}) \& (\ref{non2}--\ref{non3})),
 and in finding the complex $\omega_*$ which annihilates the determinant of the matrix derived in this way.
 This was accomplished by using the advanced linear-algebra functionalities available in MATLAB.
 It has been checked that all the results given below are free of numerical errors
 by discretising the domain by means of finer and finer 1D meshes.

\subsection{Influence of domain size, wall conditions and mesh refinement} \label{i}

\begin{figure}
 \centering
 \includegraphics[height=4.5cm]{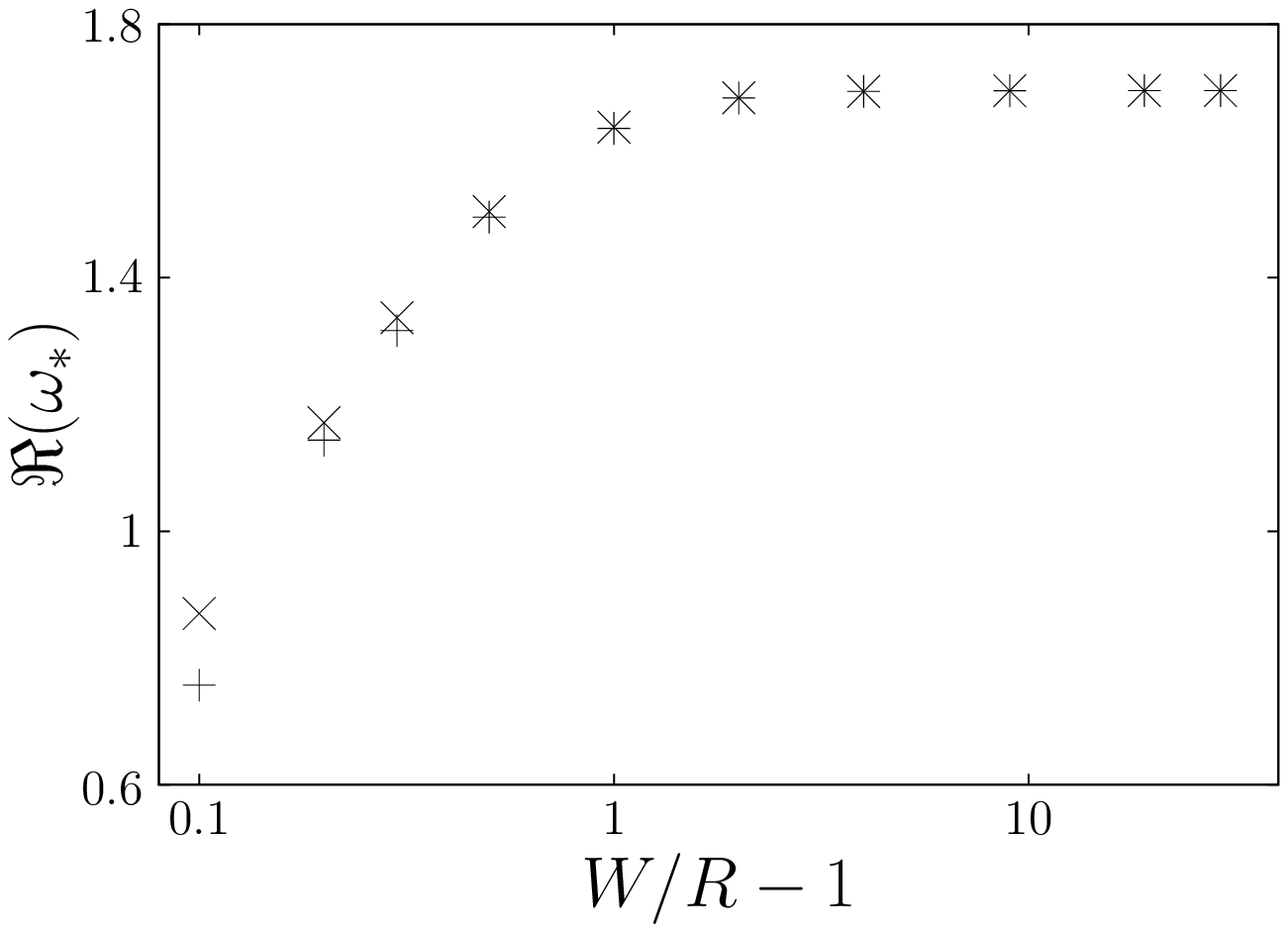}\hfill\includegraphics[height=4.5cm]{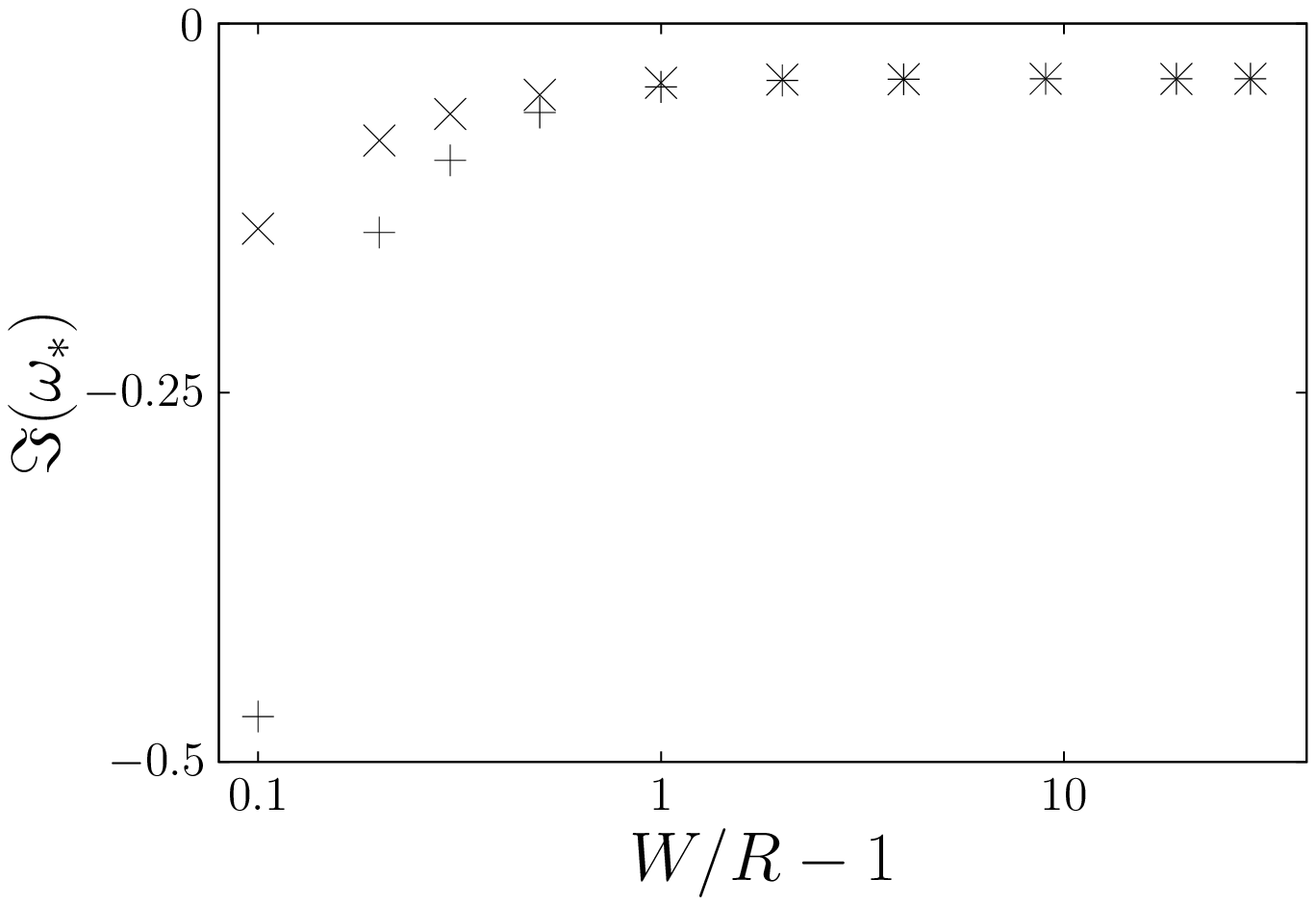}
 \caption{Behaviour of the real (left panel) and imaginary (right panel) parts
  of the complex nondimensional angular frequency $\omega_*$, as a function of the nondimensional excess wall radius $W/R-1$,
  for both the no-slip ($+$) and the free-slip ($\times$) cases. Our working point for the other calculations, $W/R-1=9$,
  lies in a range where saturation has taken place and the wall condition is irrelevant.
  The values of the other parametres are $n=2$, $\nu_*=0.001$ and $R_*=0.5$.}
 \label{W}
\end{figure}

 The presence of an external wall of circular shape has been introduced
 for the sake of simplicity, from both the analytical and computational points of view.
 However, our claim is that our results are generally valid also in the absence of this wall,
 and to state this we have to show two properties: first, that the reference case
 already belongs to a saturation zone, in which the wall is far enough as not to substantially modify
 the results if it is moved further away; second, that at this $W$ it is indifferent to impose
 no-slip or free-slip wall conditions. The latter constraint, instead of $u_{\theta}|_{r=W}=0$, translates into
 the absence of any tangential stress:
 \[0=\left.r\frac{\partial(u_{\theta}/r)}{\partial r}\right|_{r=W}+\left.\frac{1}{r}\frac{\partial u_r}{\partial\theta}\right|_{r=W}=\phi^{(2)}(W)-\frac{1}{W}\phi^{(1)}(W)\;.\]
 Figure~\ref{W} shows that these two properties are actually verified already at $W/R=10$
 (our working point for the other calculations).

\begin{figure}
 \centering
 \includegraphics[height=4.5cm]{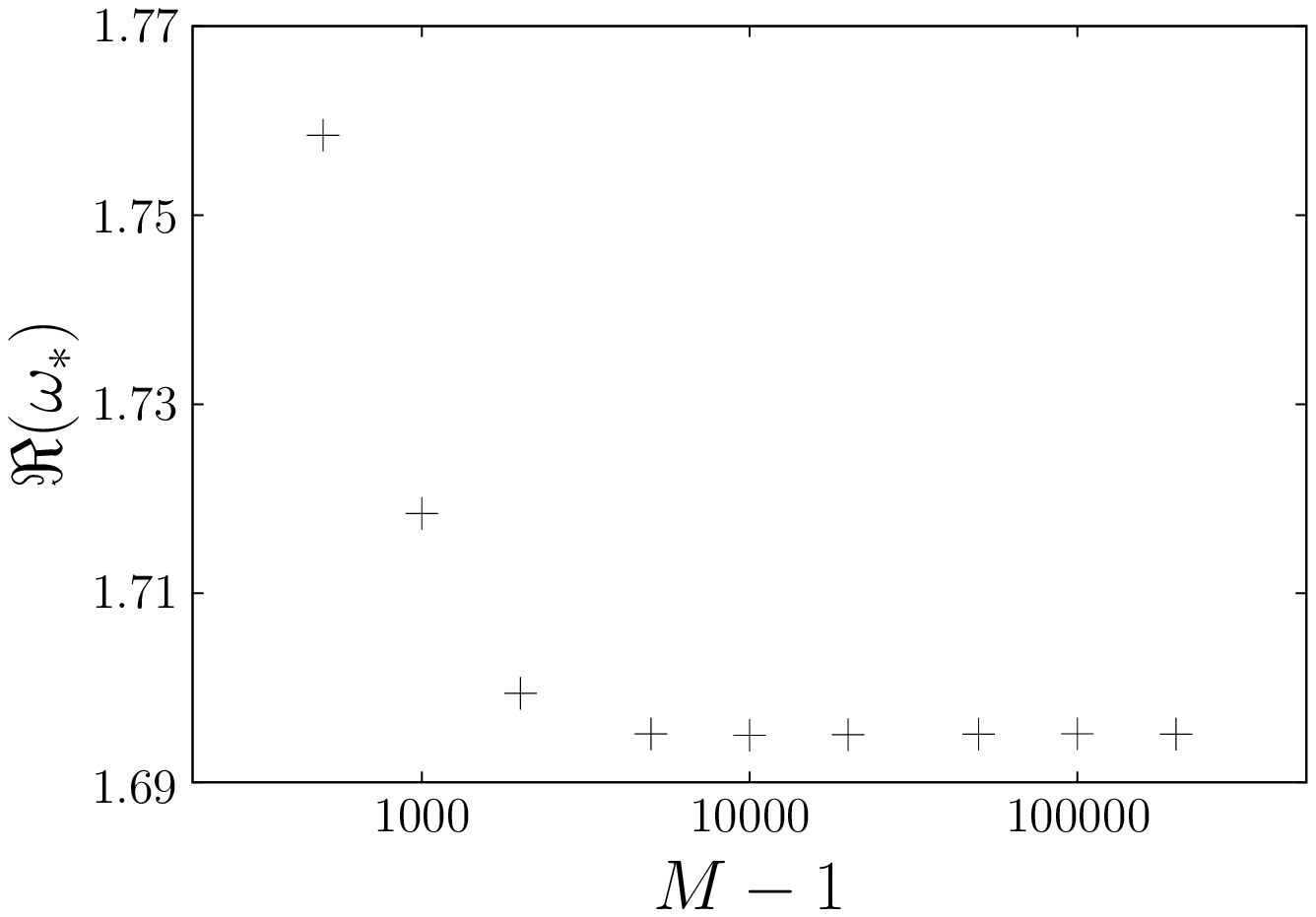}\hfill\includegraphics[height=4.5cm]{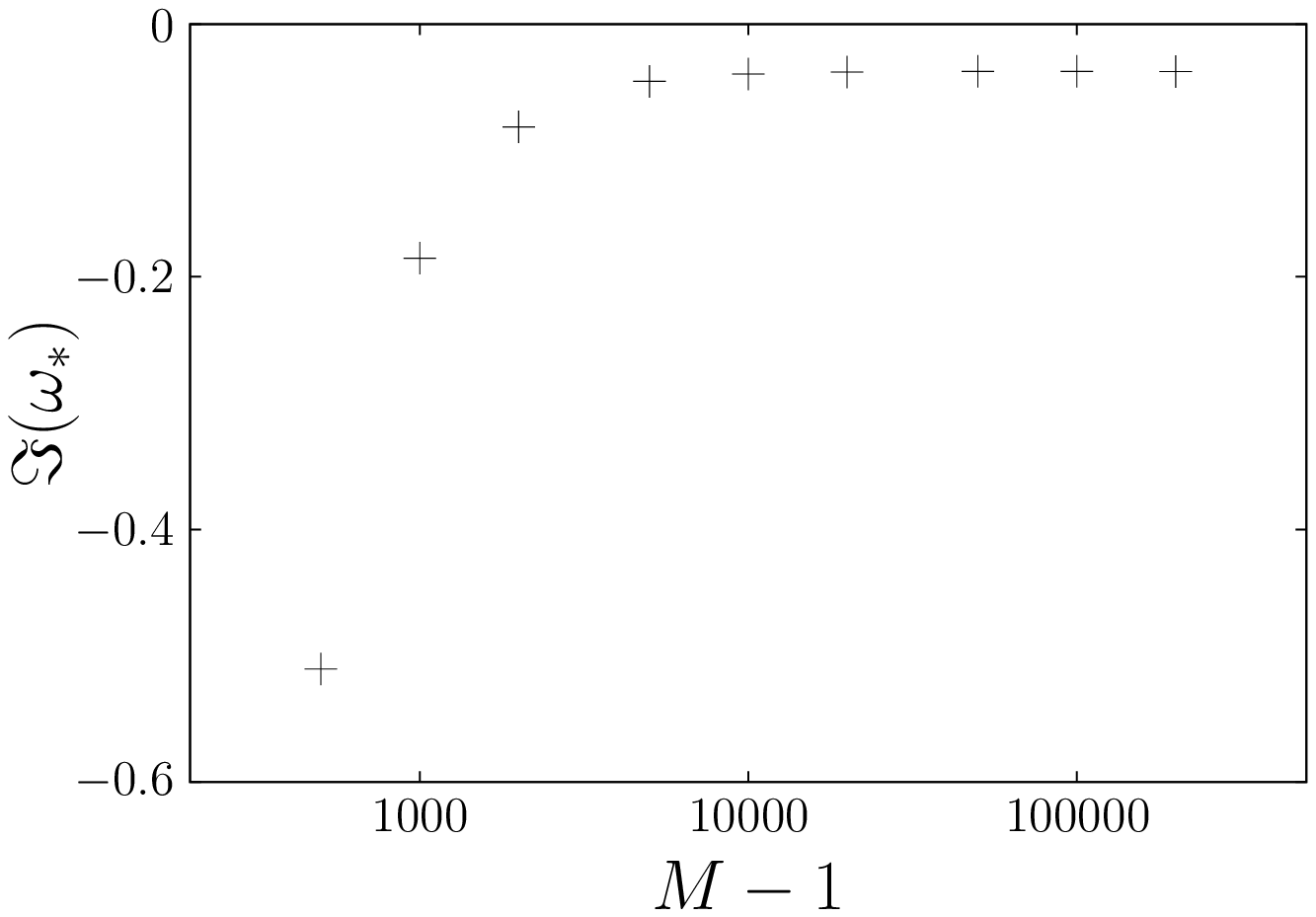}
 \caption{Same as in figure \ref{W} but as a function of $M-1$, the number of intervals in which the domain is discretised.
  Our working point for the other calculations, $M-1=100000$,
  lies in a range where convergence has taken place.}
 \label{numdeltsigm}
\end{figure}

 Finally, it is necessary to show that our numerical scheme has converged,
 in the sense that a reasonable change in the mesh size should not alter the results substantially.
 Figure \ref{numdeltsigm} shows indeed the behaviour of the complex frequency versus the number of discretisation points $M$:
 no appreciable departure from the convergence is noticeable at our working point $M=100001$.

\subsection{Mode frequencies} \label{ii}

\begin{figure}
 \centering
 \includegraphics[height=4.5cm]{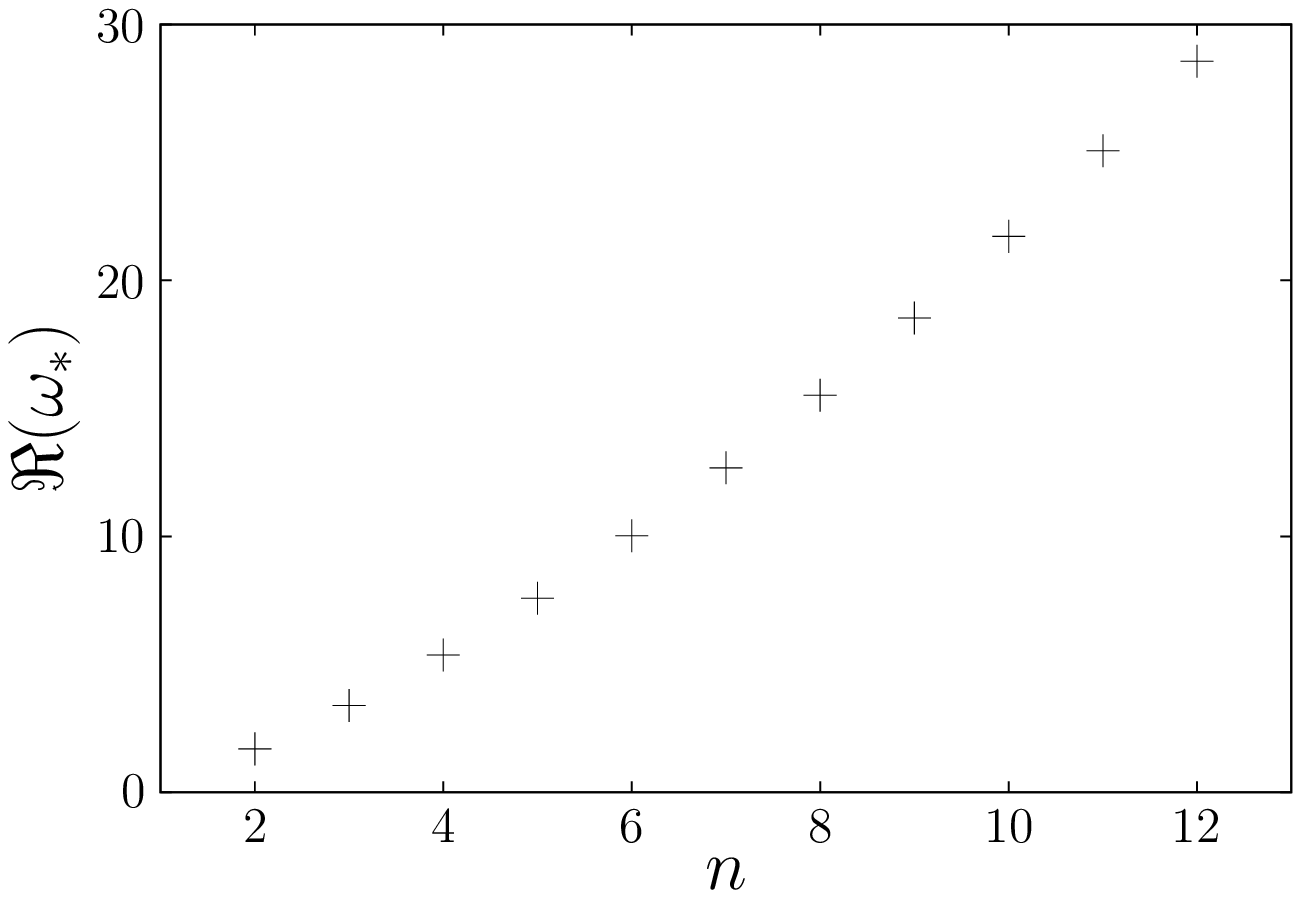}\hfill\includegraphics[height=4.5cm]{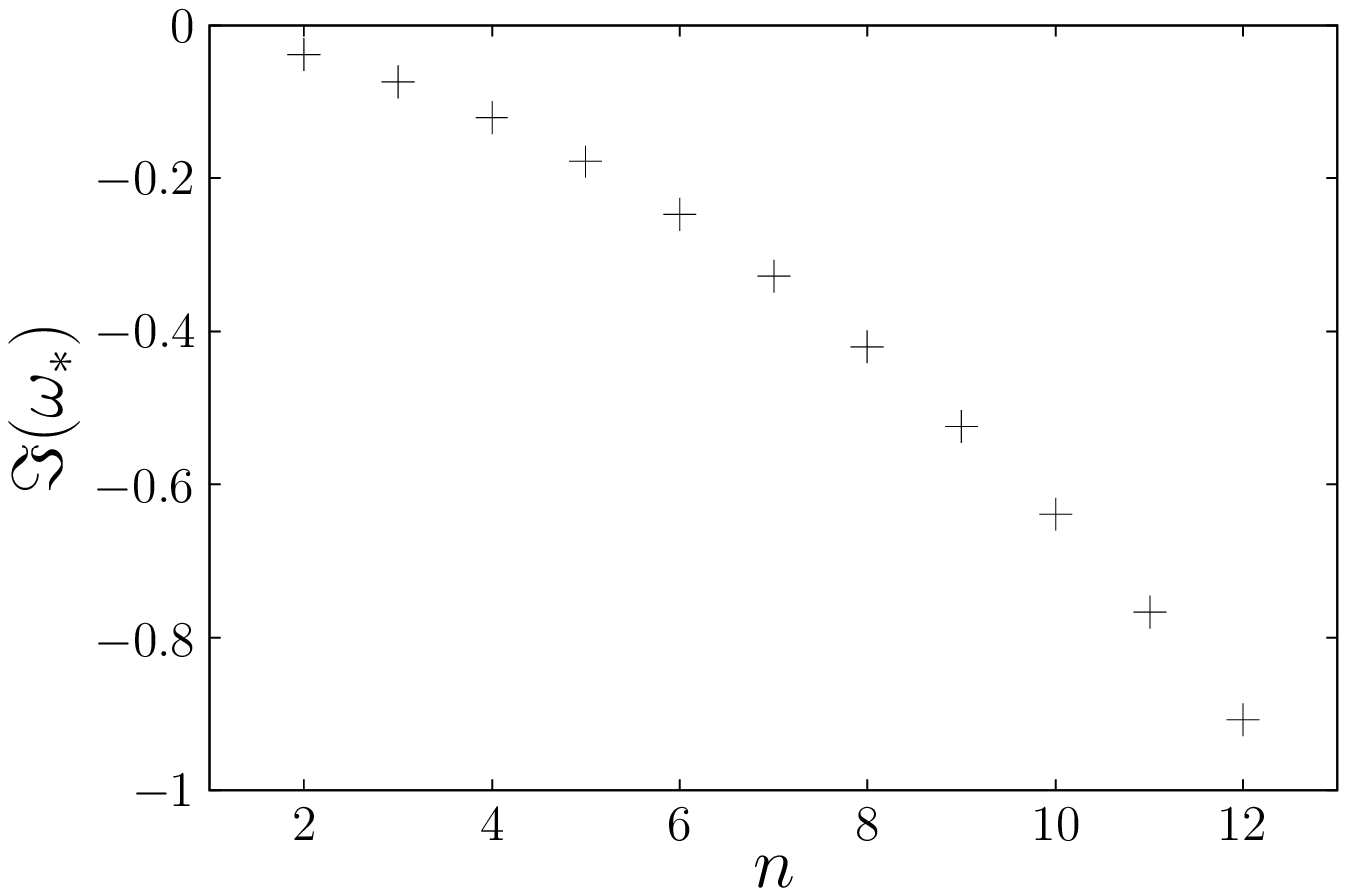}
 \caption{Same as in figure \ref{W} but as a function of the oscillation mode $n$,
  for fixed values of $\nu_*=0.001$ and $R_*=0.5$.
  The modes $n=0$ and $n=\pm1$ must be neglected because they would imply
  an angle-independent evolution and a translation of the center of mass, respectively.}
 \label{enne}
\end{figure}

 Figure \ref{enne} shows the dependence of $\omega_*$ on the oscillation mode $n$.
 An increase in both the angular frequency and the damping rate can be observed for growing $n$.

\begin{figure}
 \centering
 \includegraphics[height=4.5cm]{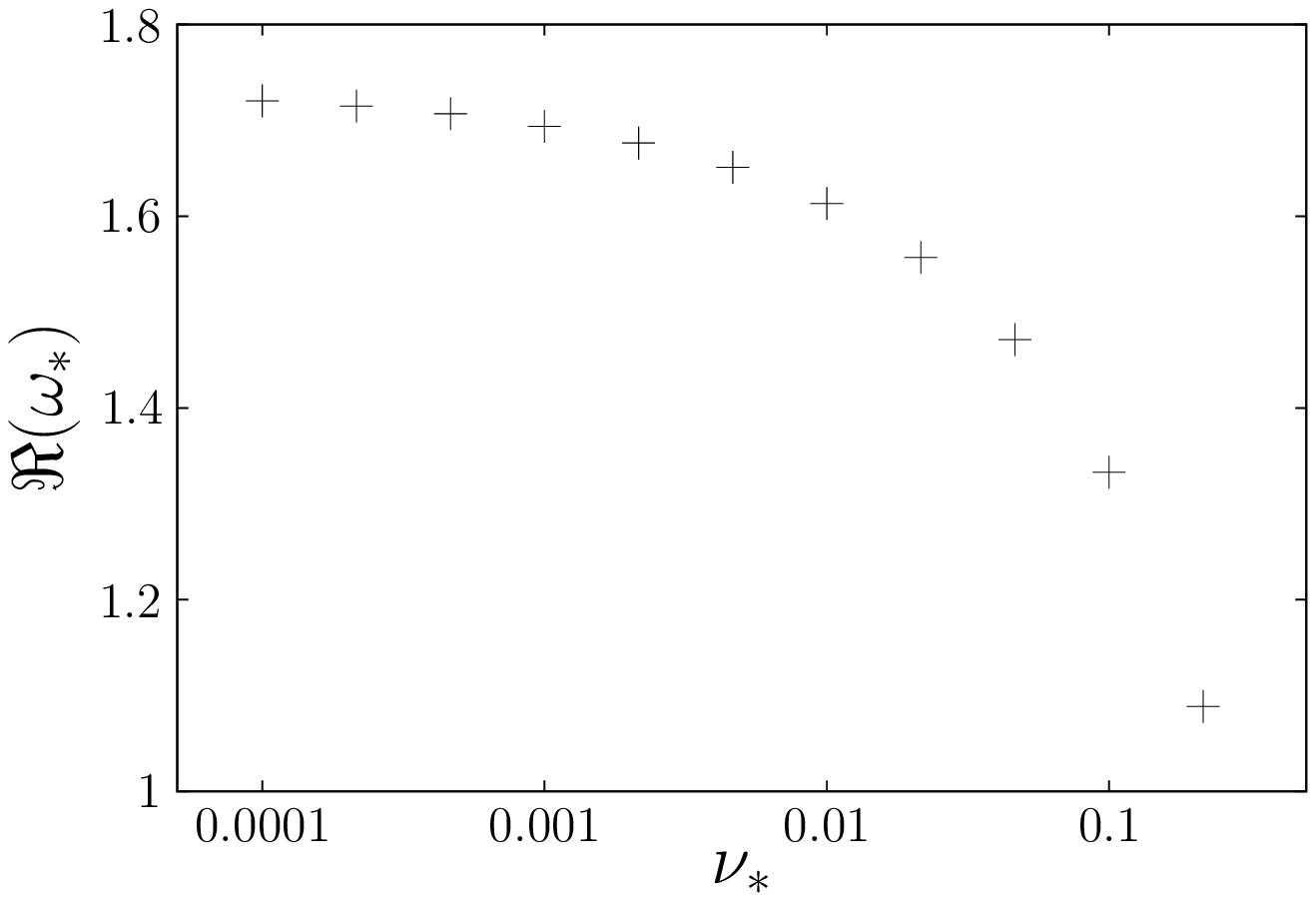}\hfill\includegraphics[height=4.5cm]{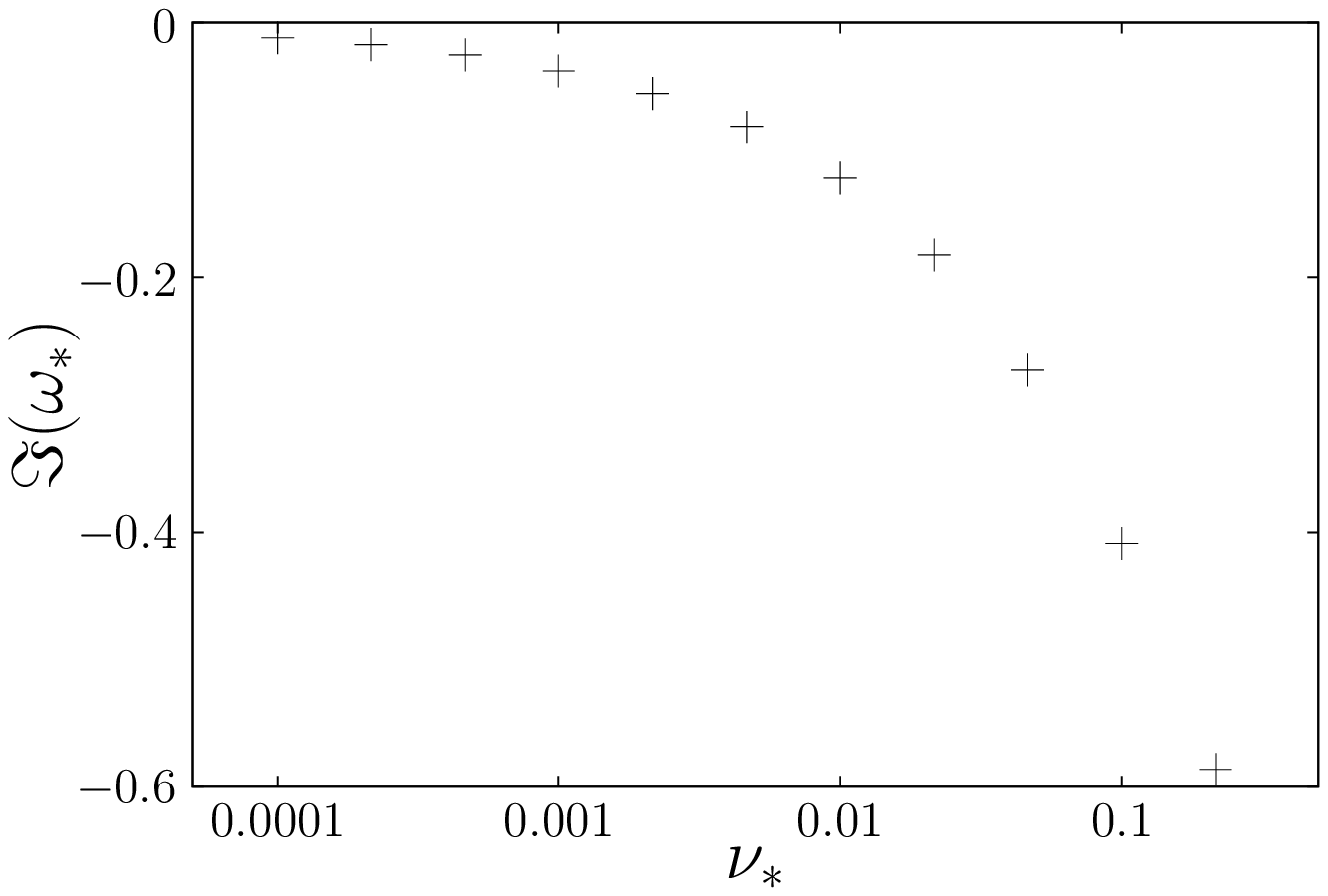}
 \caption{Same as in figure \ref{W} but as a function of the reduced viscosity $\nu_*=\nu T/R^2$,
  for fixed values of $n=2$ and $R_*=0.5$.}
 \label{ni}
\end{figure}

 Let us now focus on the mode $n=2$, which is the closest to an ellipse and the most used in numerical studies
 \citep{Tu:1992,Lee:2003,Cottet:2006,Song:2008,Tan:2008},
 and let us investigate the dependence of $\omega_*$ on the reduced viscosity $\nu_*$.
 As shown in figure \ref{ni}, a decrease in $\nu_*$ leads to an increase of the real part $h$,
 and to a decrease of the absolute value of the imaginary part $g$.
 Notice that this latter tends to zero in the limit of vanishing fluid viscosity,
 because in the absence of any imposed membrane dissipation every damping effect disappears there.

\begin{figure}
 \centering
 \includegraphics[height=4.35cm]{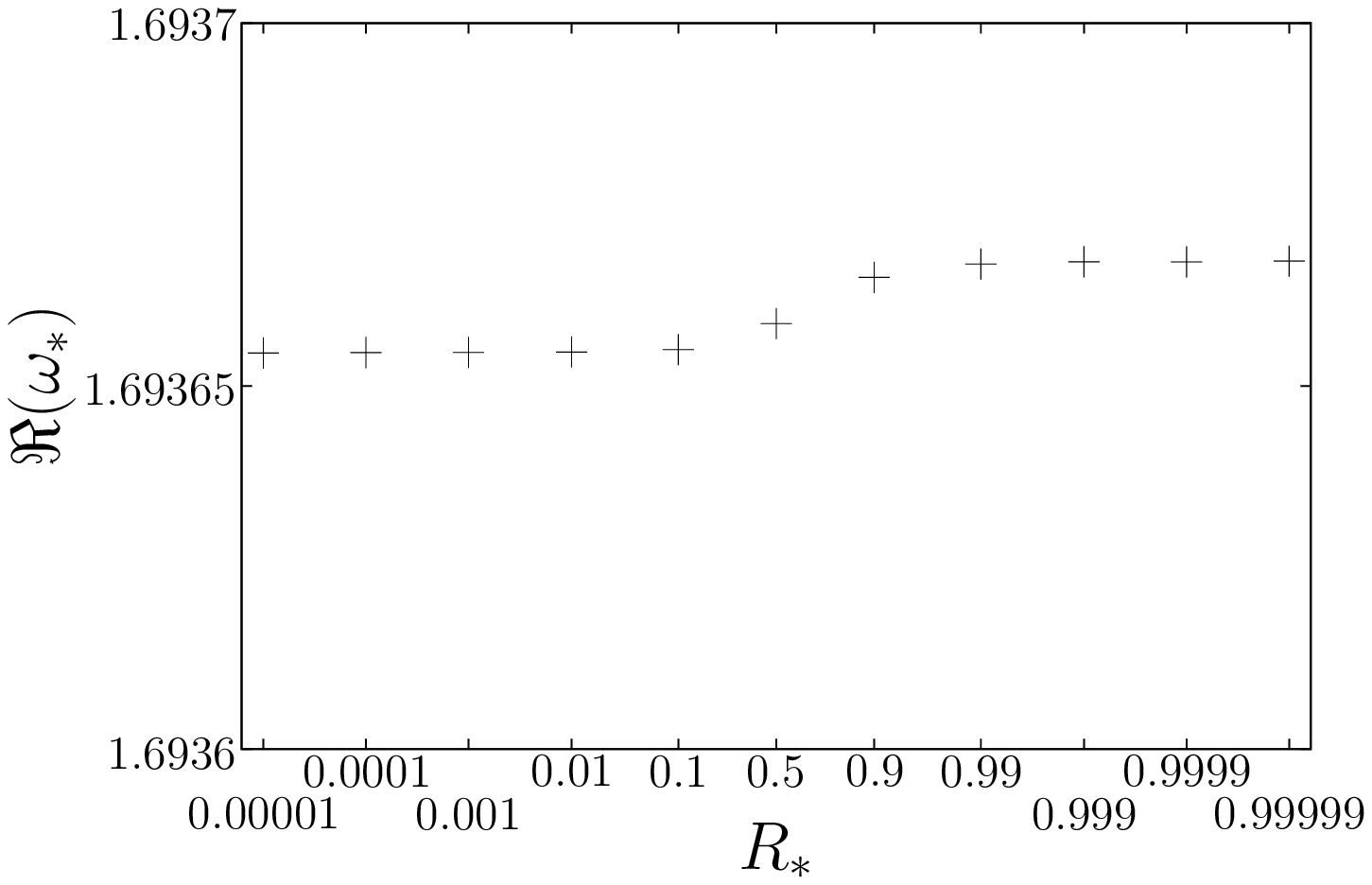}\hfill\includegraphics[height=4.35cm]{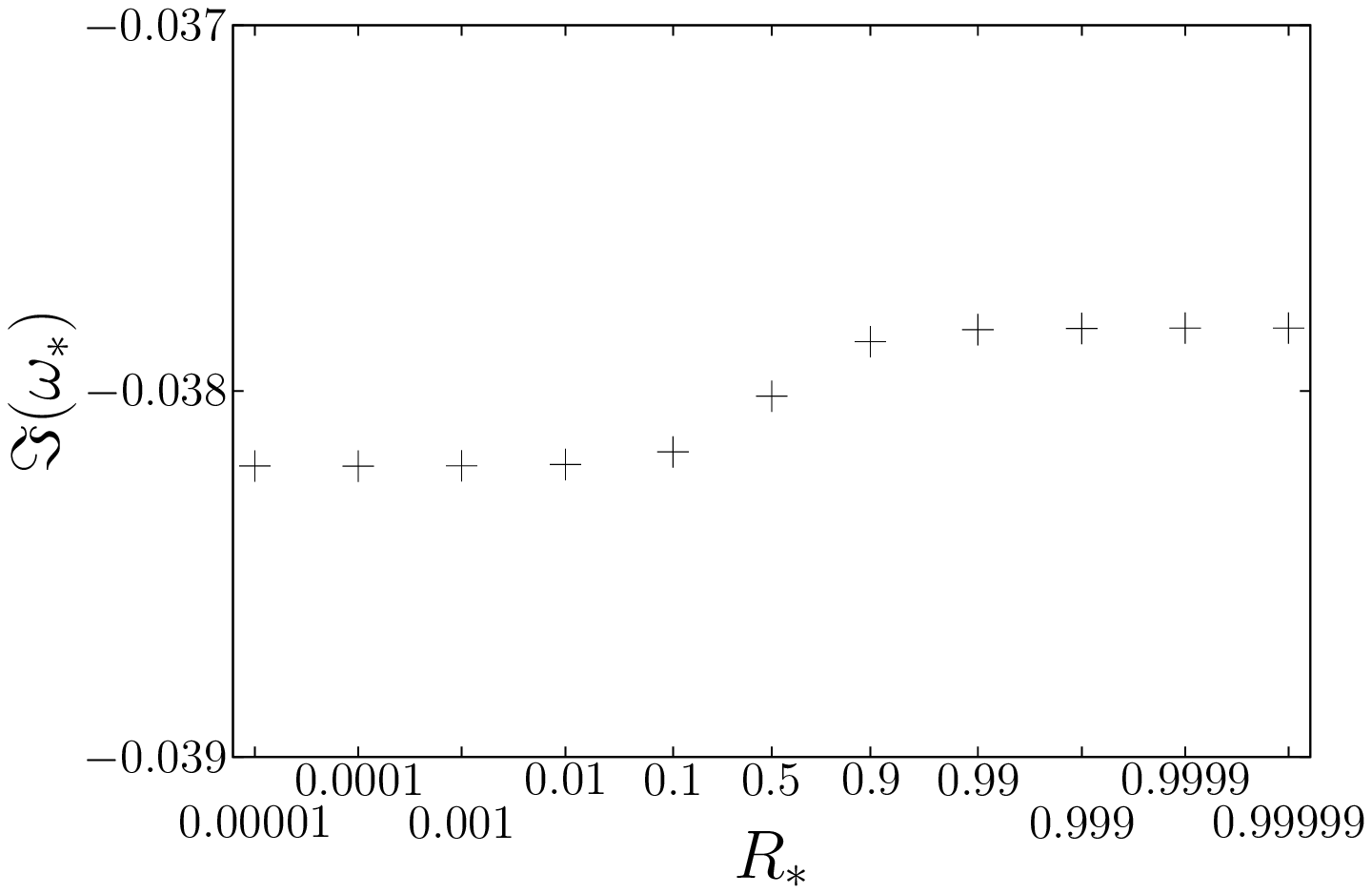}
 \caption{Same as in figure \ref{W} but as a function of the ratio $R_*= R_0/R$,
  for fixed values of $n=2$ and $\nu_*=0.001$. Notice the uncharacteristic scales for both axes:
  ordinates are plotted in a very expanded scale (i.e.\ with a very reduced range)
  to be able to notice any variation; abscissae are represented in a scale linear in the variable
  $\ln[R_*/(1-R_*)]$ in order to show the behaviour in the limits $R_*\to0_+$ and $R_*\to1_-$.}
 \label{erre}
\end{figure}

 For what concerns the dependence of $\omega_*$ on the remaining parametre $R_*$,
 we found that it is very feeble, which is shown in figure \ref{erre}.
 This is due to the fact that the core of the dependence of $\omega$ on $R$ and $R_0$
 has already been properly taken into account through the definitions of the membrane
 time scale $T$ and of the reduced viscosity (\ref{T}--\ref{Sm}), where the block
 $E(R-R_0)/R_0$ can be identified as a surface tension. Consequently,
 the dependence on $R_*$ has disappeared in (\ref{nondi}) and (\ref{non3}),
 and only lies in the jump of the second derivative (\ref{non2}).

 Notice that the aforementioned dependences on $n$, $\nu_*$ and $R_*$
 make it impossible to find two (or more) distinct situations,
 namely with different oscillations modes, such as to have the same complex $\omega_*$.
 This is due to the monotonically-growing character of both $h$ and $|g|$ in $n$,
 which --- taking into account the marginal role played by $R_*$ ---
 cannot be compensated by a variation in $\nu_*$, as e.g.\ a reduction of the latter
 would imply a decrease in $\Re(\omega_*)$ but an increase in $|\Im(\omega_*)|$.
 However, this conclusion is no longer true for the dimensional complex $\omega$,
 which can be given the same value in different modes by appropriately tuning the other parametres.

\subsection{Spatial structure of modes} \label{iii}

\begin{figure}
 \centering\vspace{0.5cm}
 \includegraphics[height=5cm]{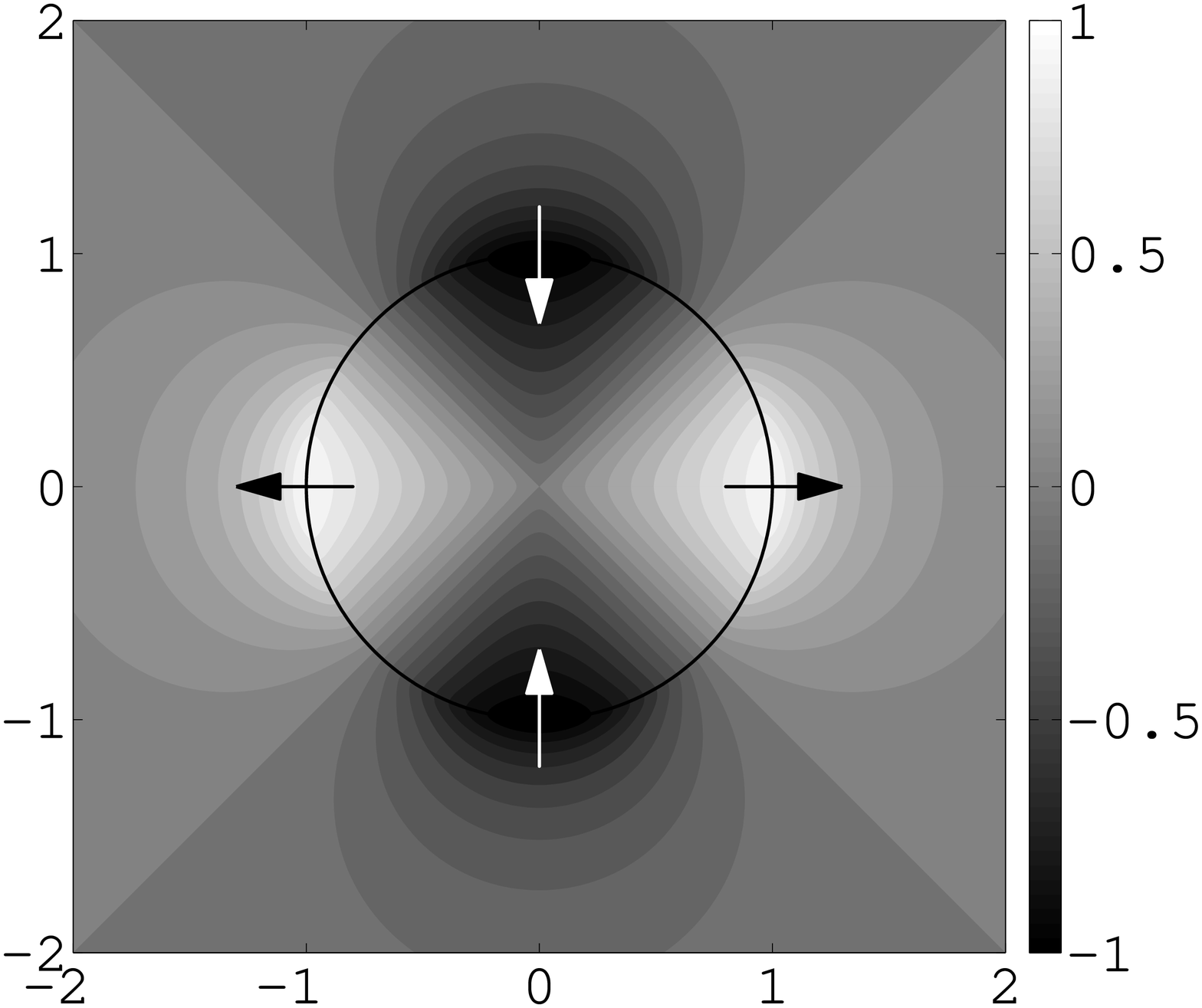}\hfill\includegraphics[height=5cm]{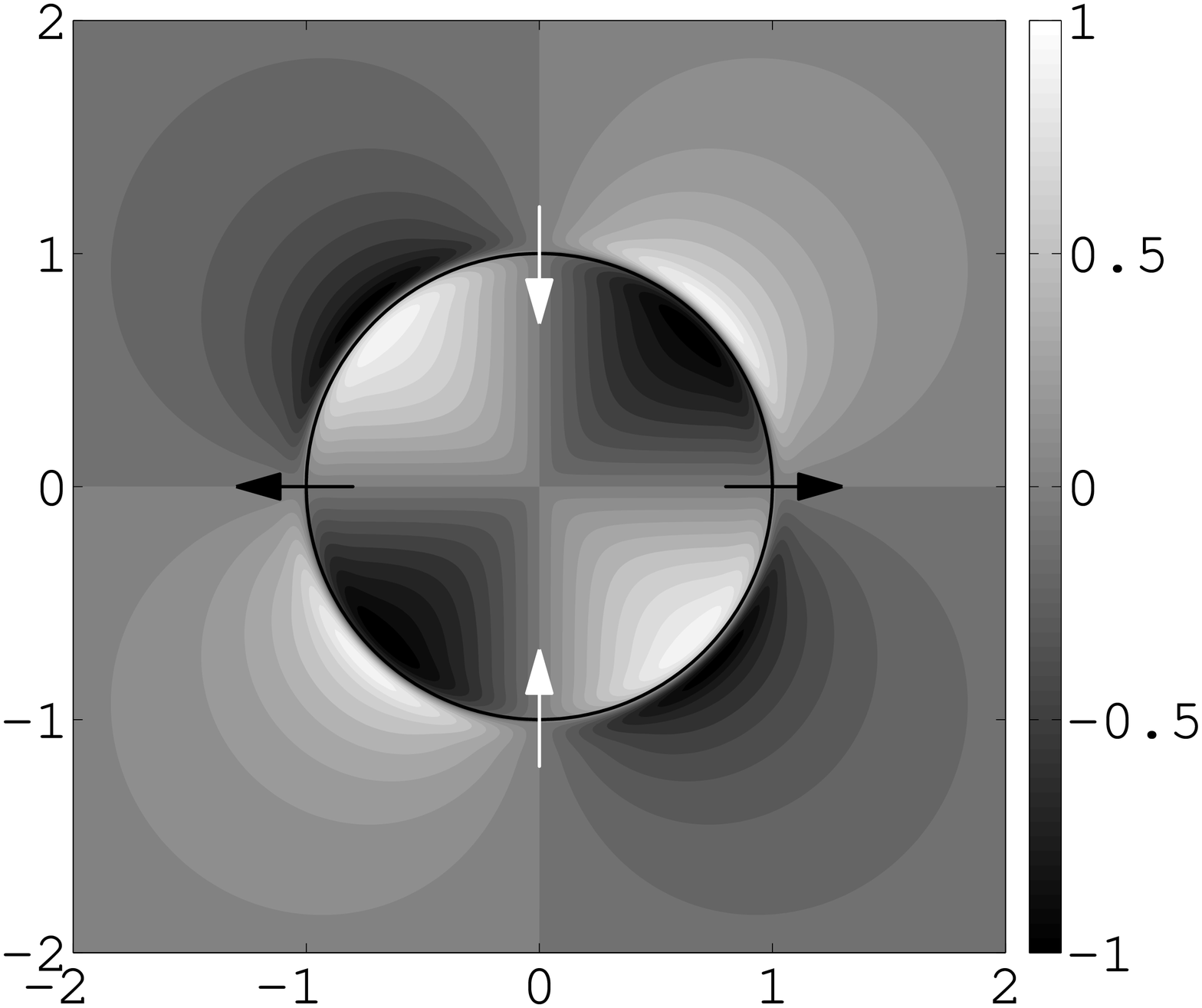}\\[0.5cm]
 \includegraphics[height=5cm]{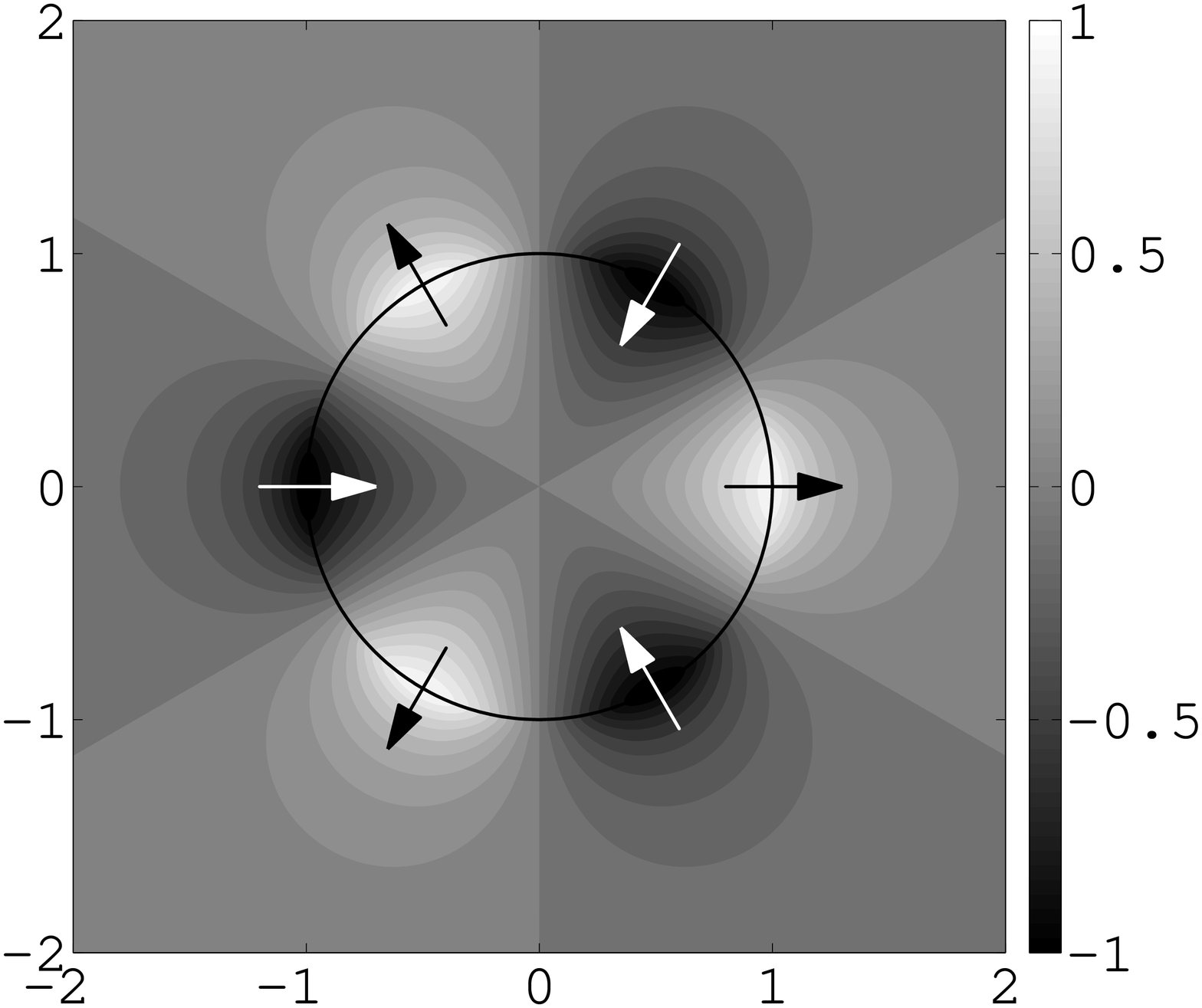}\hfill\includegraphics[height=5cm]{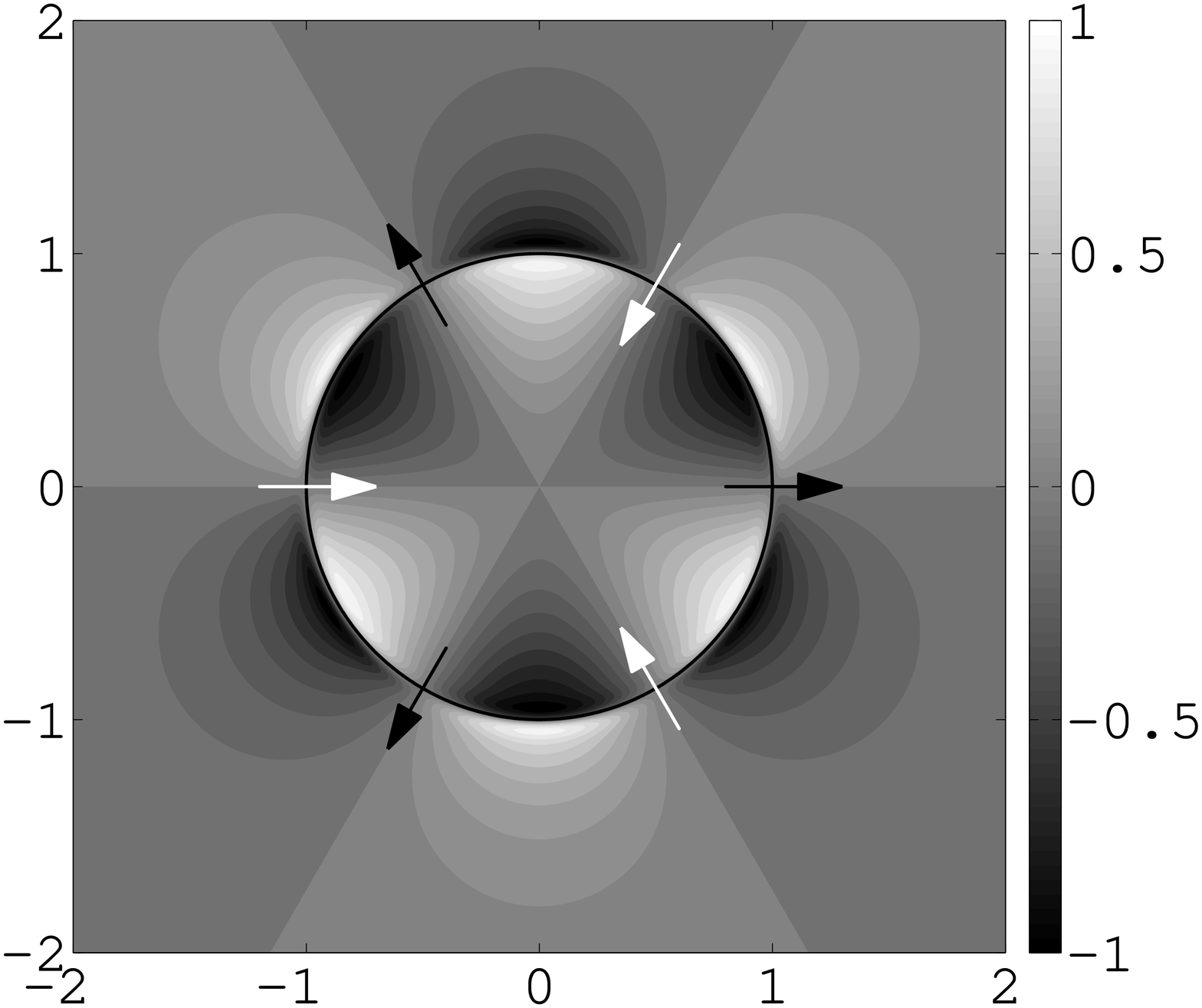}\\[0.5cm]
 \includegraphics[height=5cm]{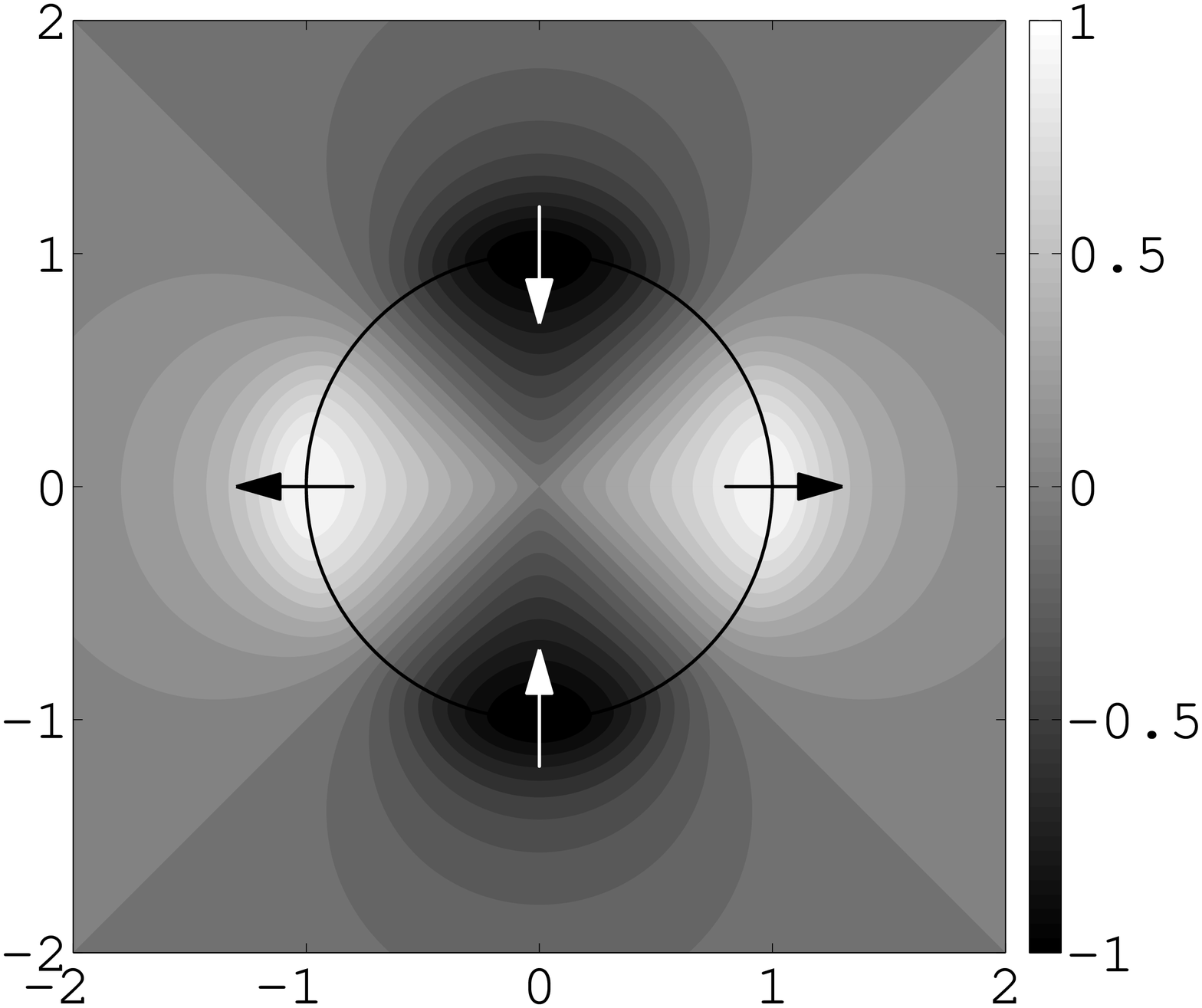}\hfill\includegraphics[height=5cm]{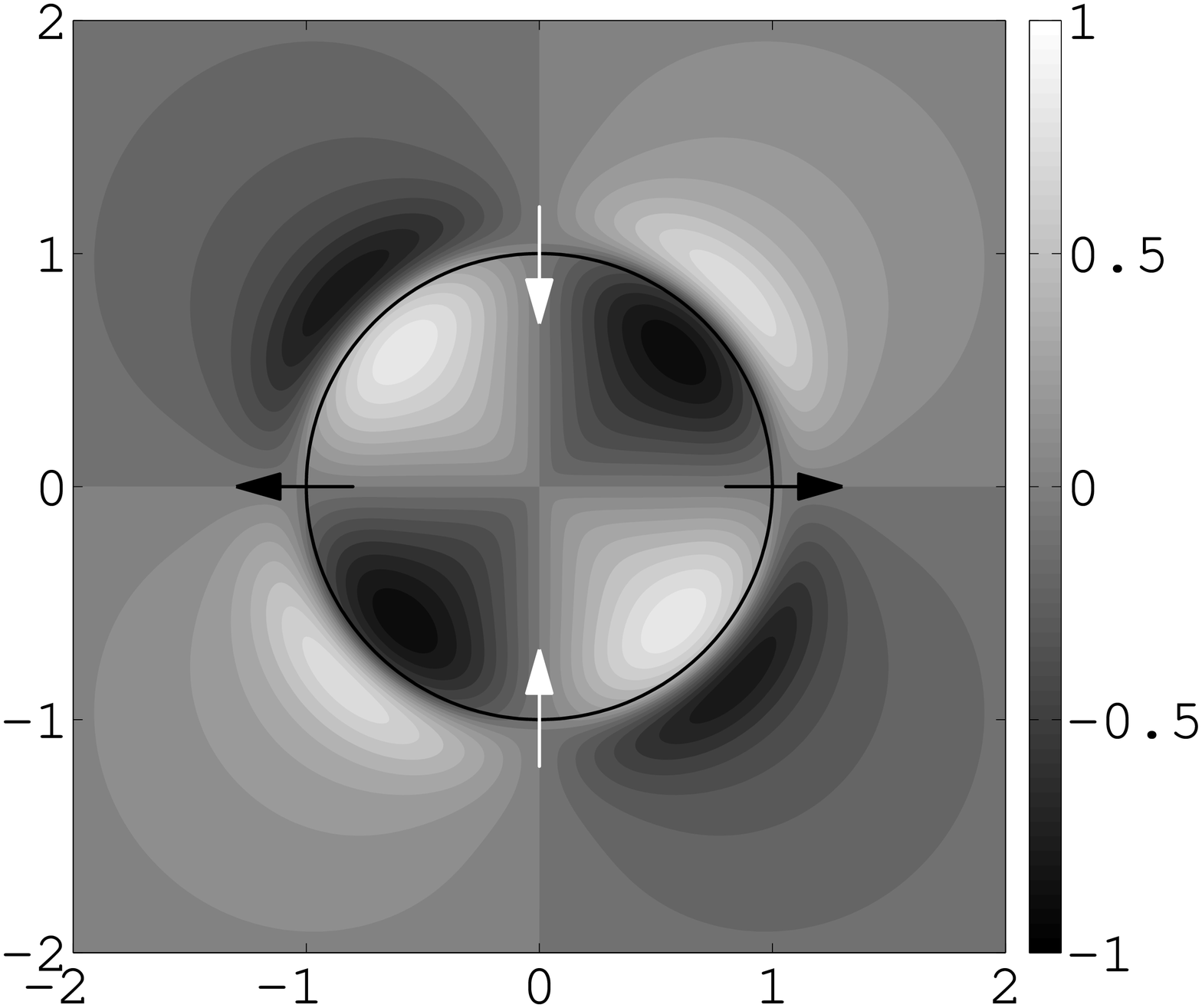}
 \caption{Normalised fluid velocity in an area around the membrane,
  captured at a passage time of the membrane through the reference circle
  (represented by the thin black line).
  Plotted are the ``standard'' case $n=2$ \& $\nu_*=0.001$ \& $R_*=0.5$ (top),
  the case with different $n=3$ (middle), and the case with increased $\nu_*=0.01$ (bottom).
  Left panels represent the radial component, right panels the azimuthal component.
  In these latter, notice that $u_{\theta}$ is continuous according to (\ref{jump}),
  but it varies very rapidly across the membrane at the angle $\theta=\upi/2n$ and odd multiples.
  The arrows denote the direction of the local instantaneous membrane motion
  and are conventionally placed at the angle $\theta=\upi/n$ and multiples,
  where $\bm{u}=u_r\hat{\bm{e}}_r$ gets its maximum absolute value $U$.
  Both velocity components are normalised with such a value $U$.}
 \label{sketch}
\end{figure}

\begin{figure}
 \centering\vspace{0.5cm}
 \includegraphics[height=5cm]{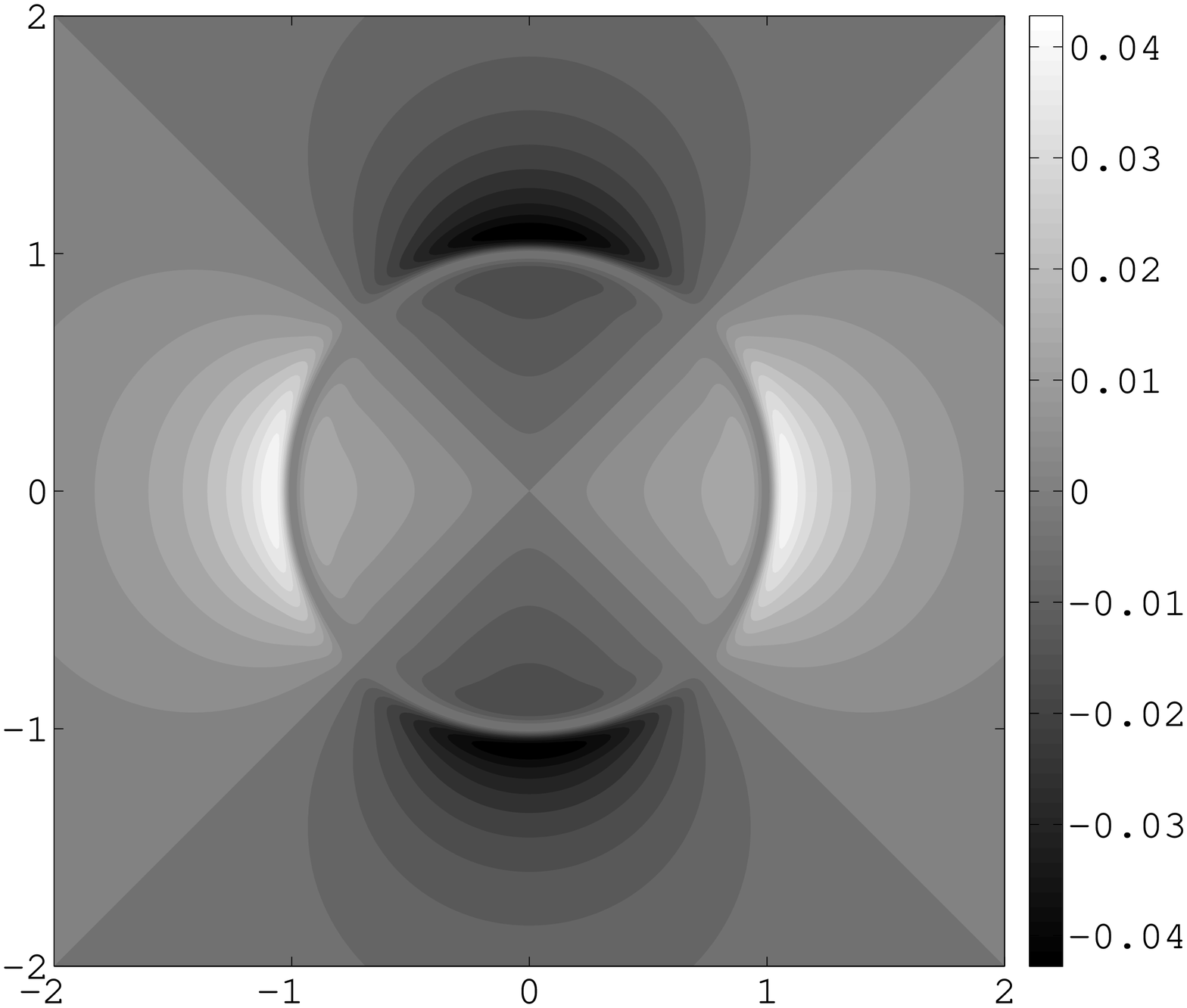}\hfill\includegraphics[height=5cm]{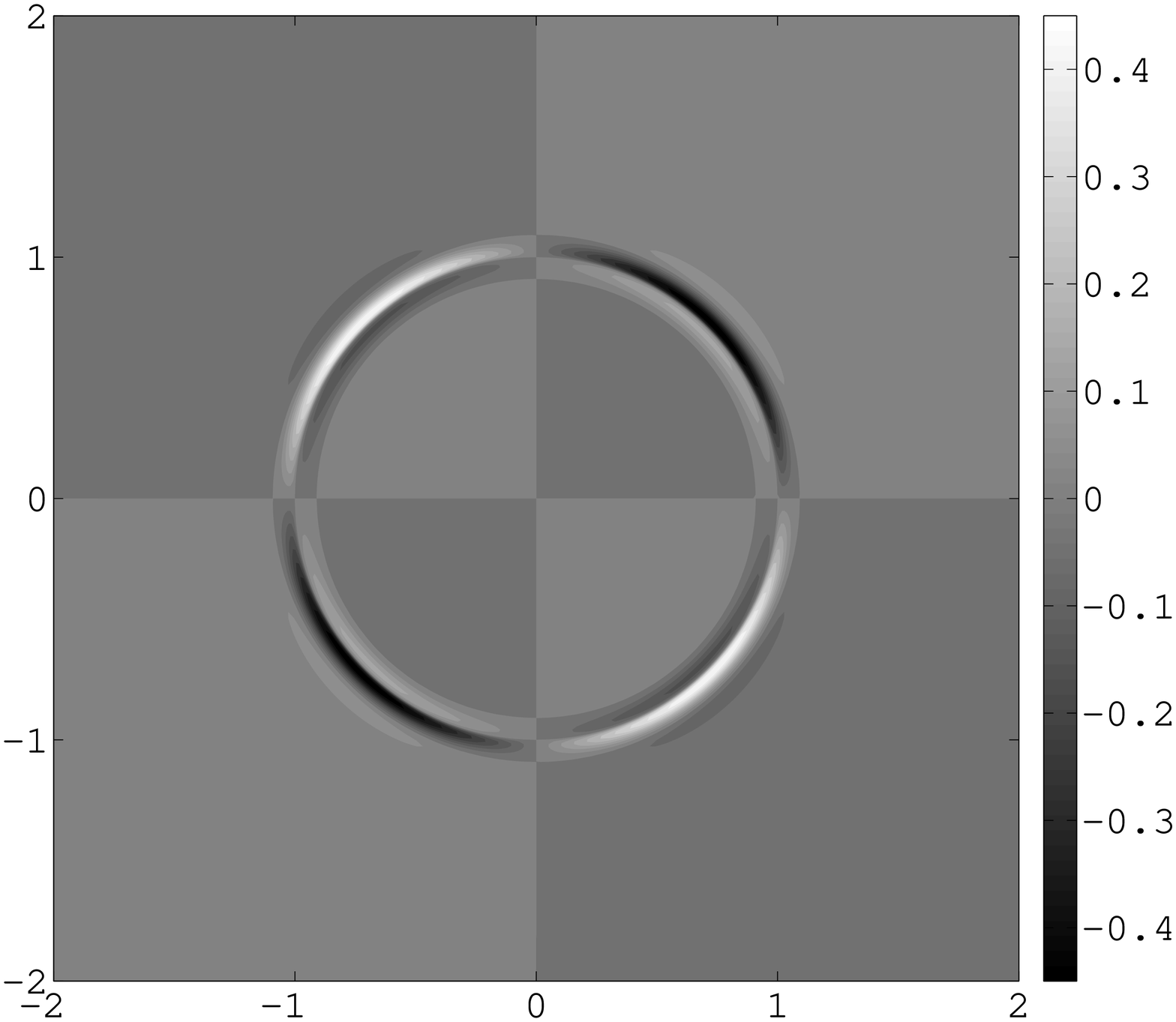}\\[0.5cm]
 \includegraphics[height=5cm]{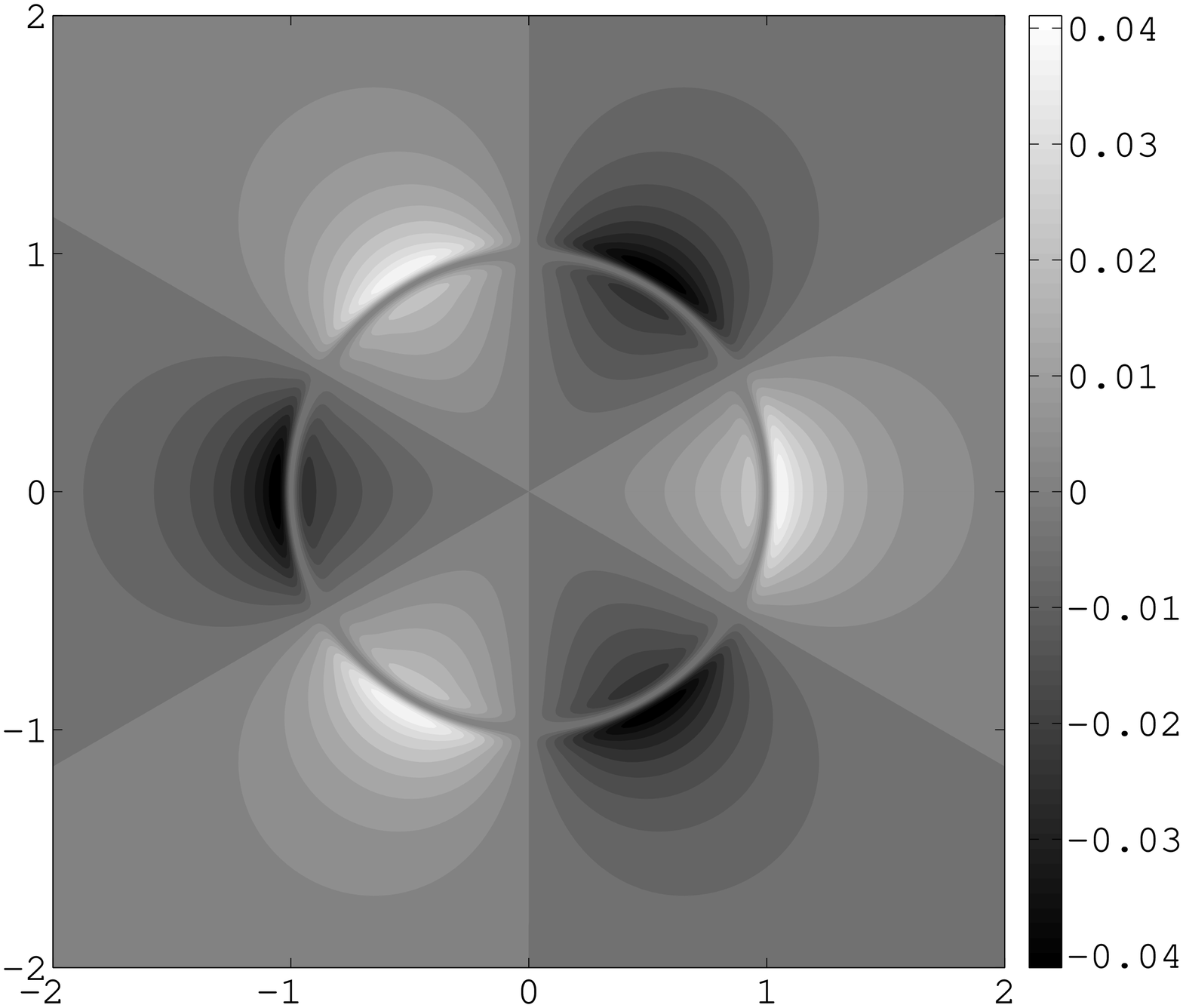}\hfill\includegraphics[height=5cm]{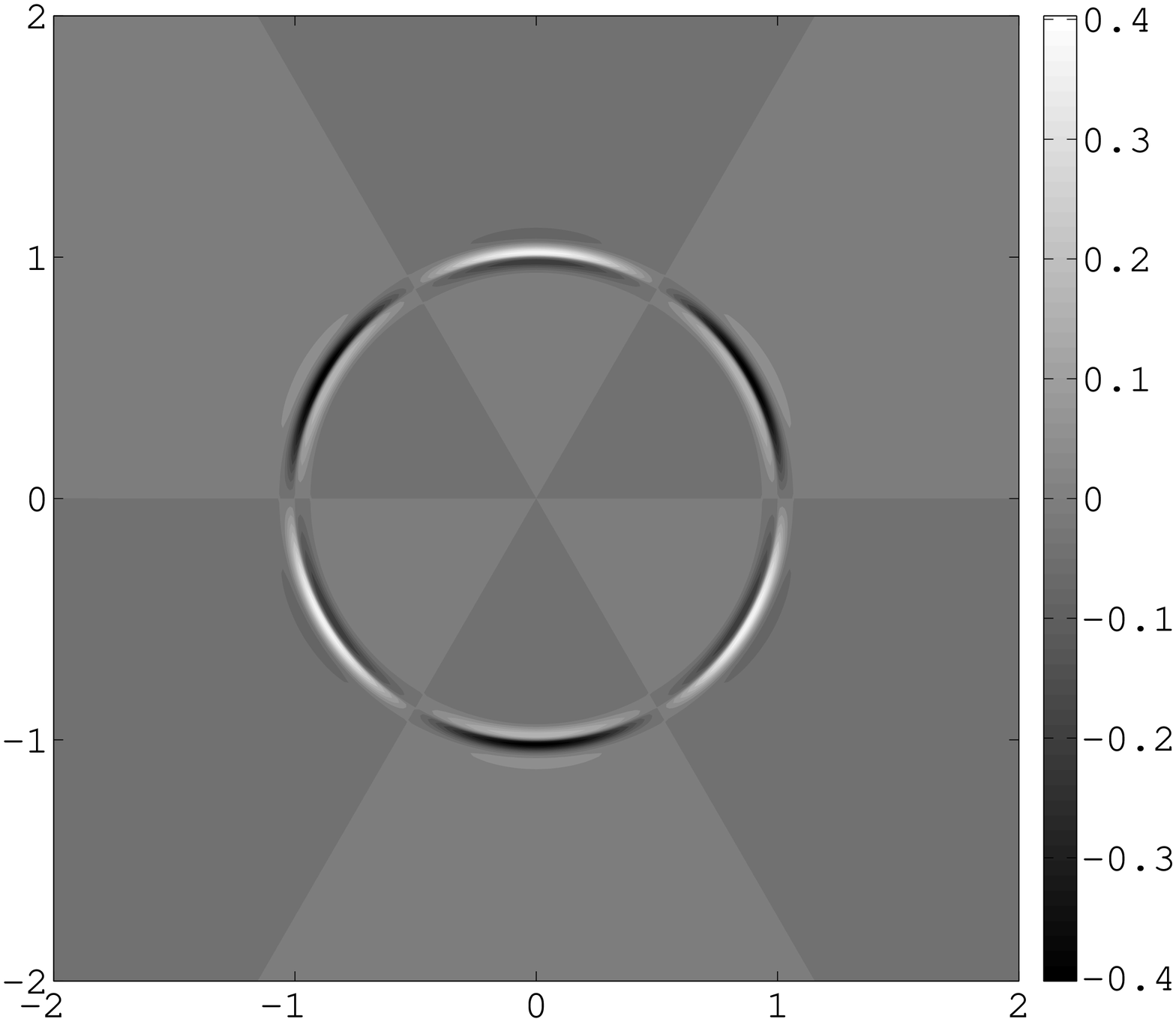}\\[0.5cm]
 \includegraphics[height=5cm]{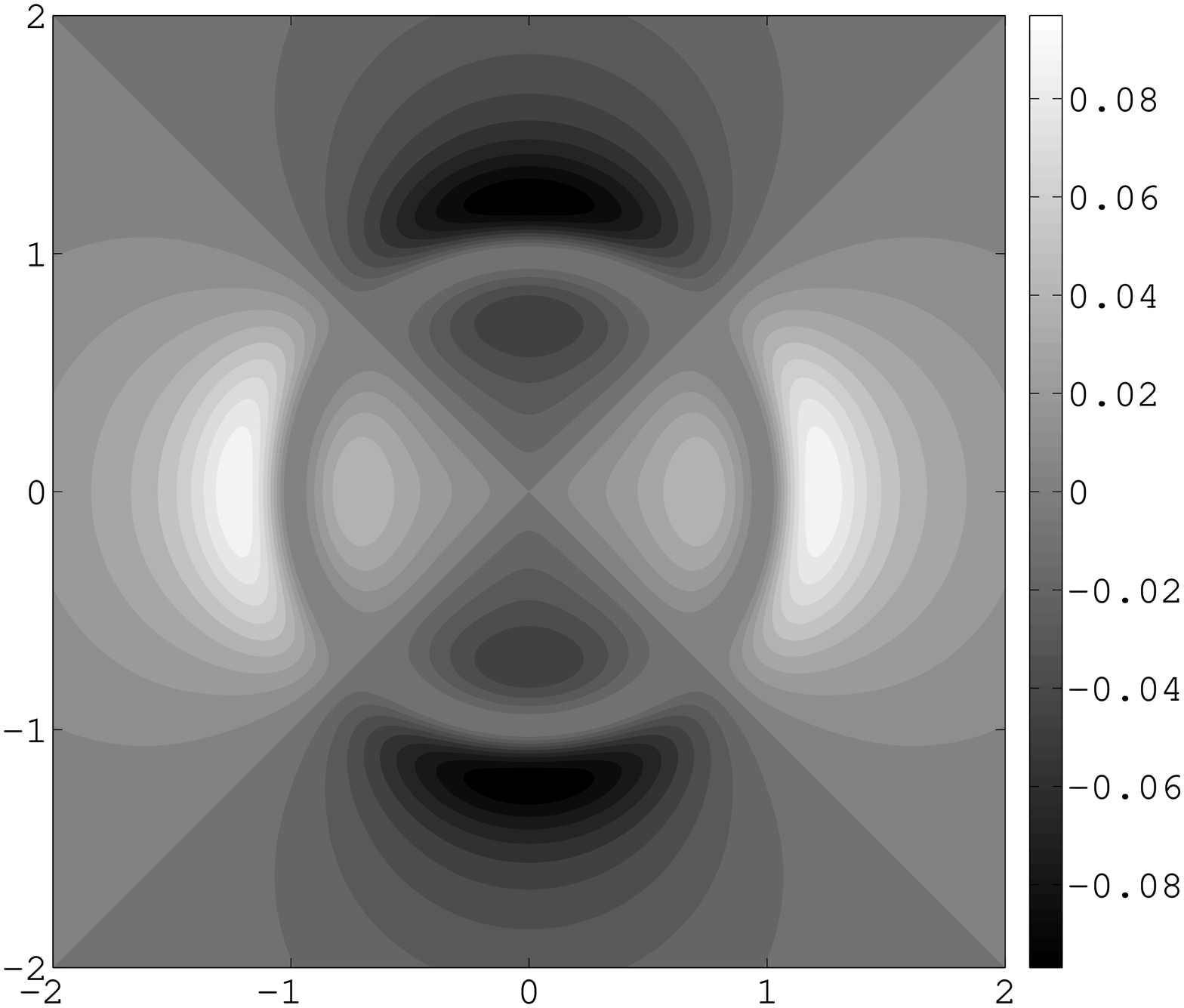}\hfill\includegraphics[height=5cm]{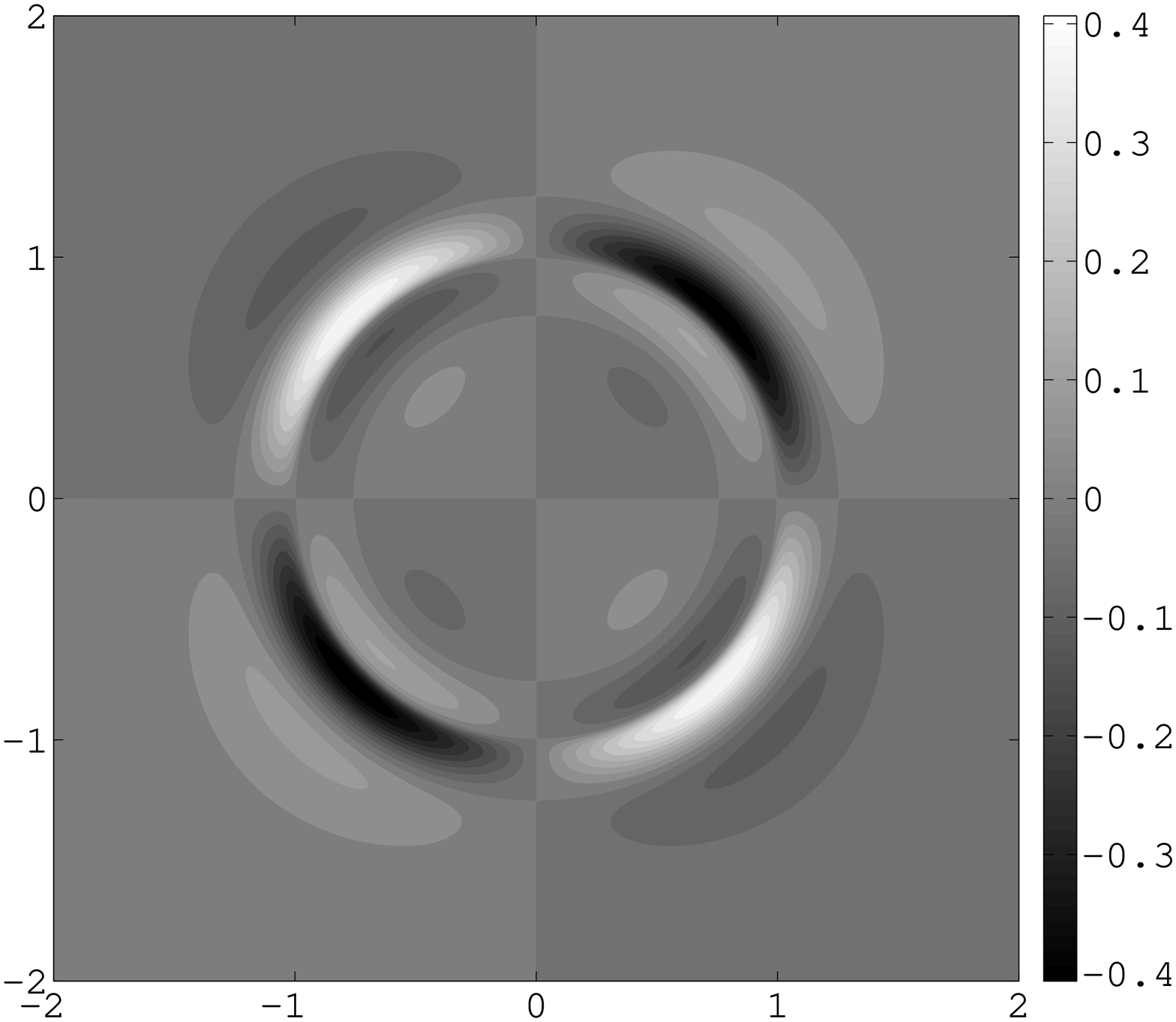}
 \caption{Same as in figure \ref{sketch} (except for the obvious absence of reference circle and arrows),
  but captured at the time of maximum membrane eccentricity when the whole membrane is at rest.
  As the normalization of both velocity components with the same above-defined $U$
  --- the maximum absolute speed value reached during the passage through the circle in figure \ref{sketch} ---
  is maintained, notice the orders of magnitude: small for the azimuthal component and even smaller for the radial one.}
 \label{elli}
\end{figure}

\begin{figure}
 \centering\vspace{0.5cm}
 \includegraphics[height=4.5cm]{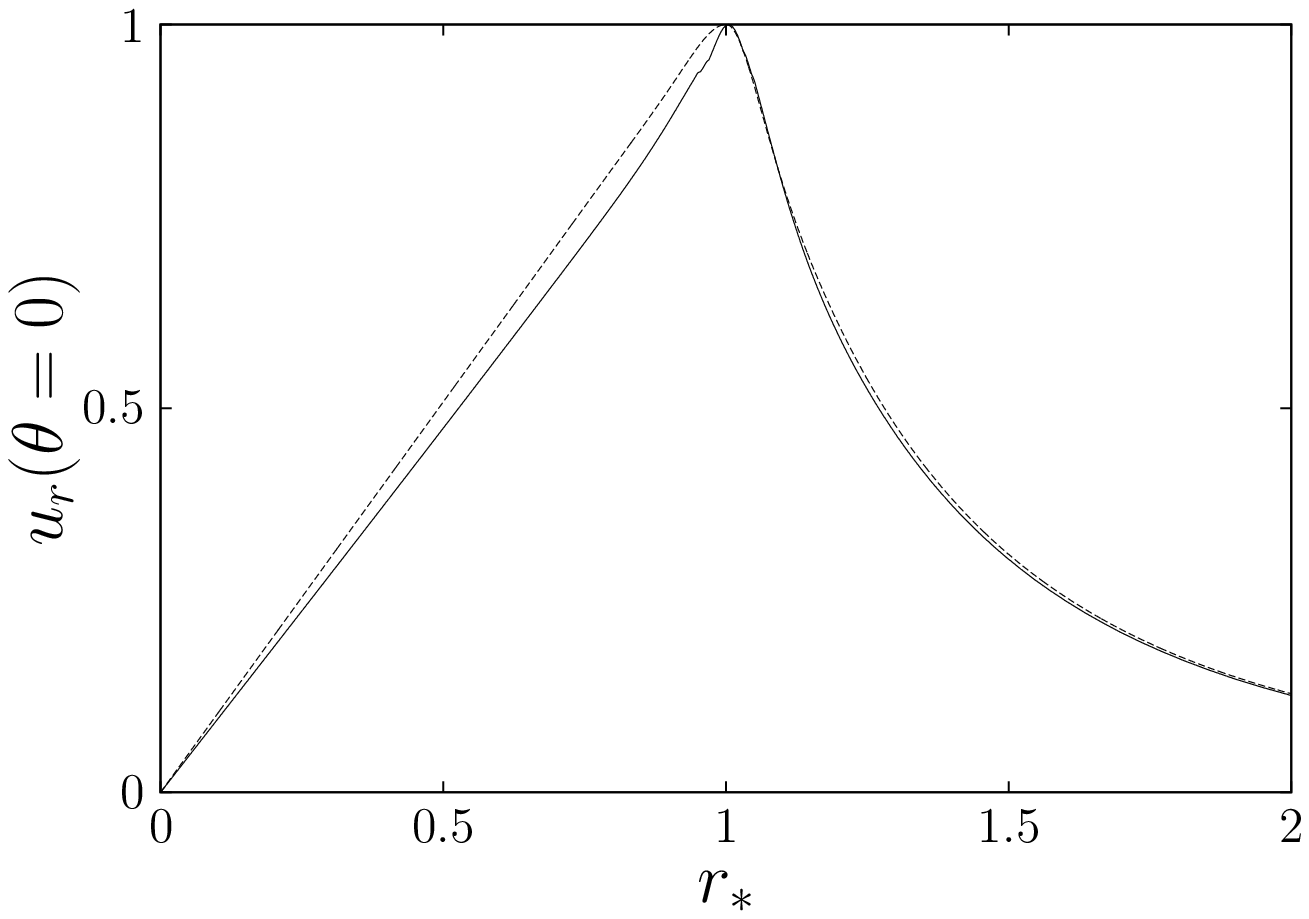}\hfill\includegraphics[height=4.5cm]{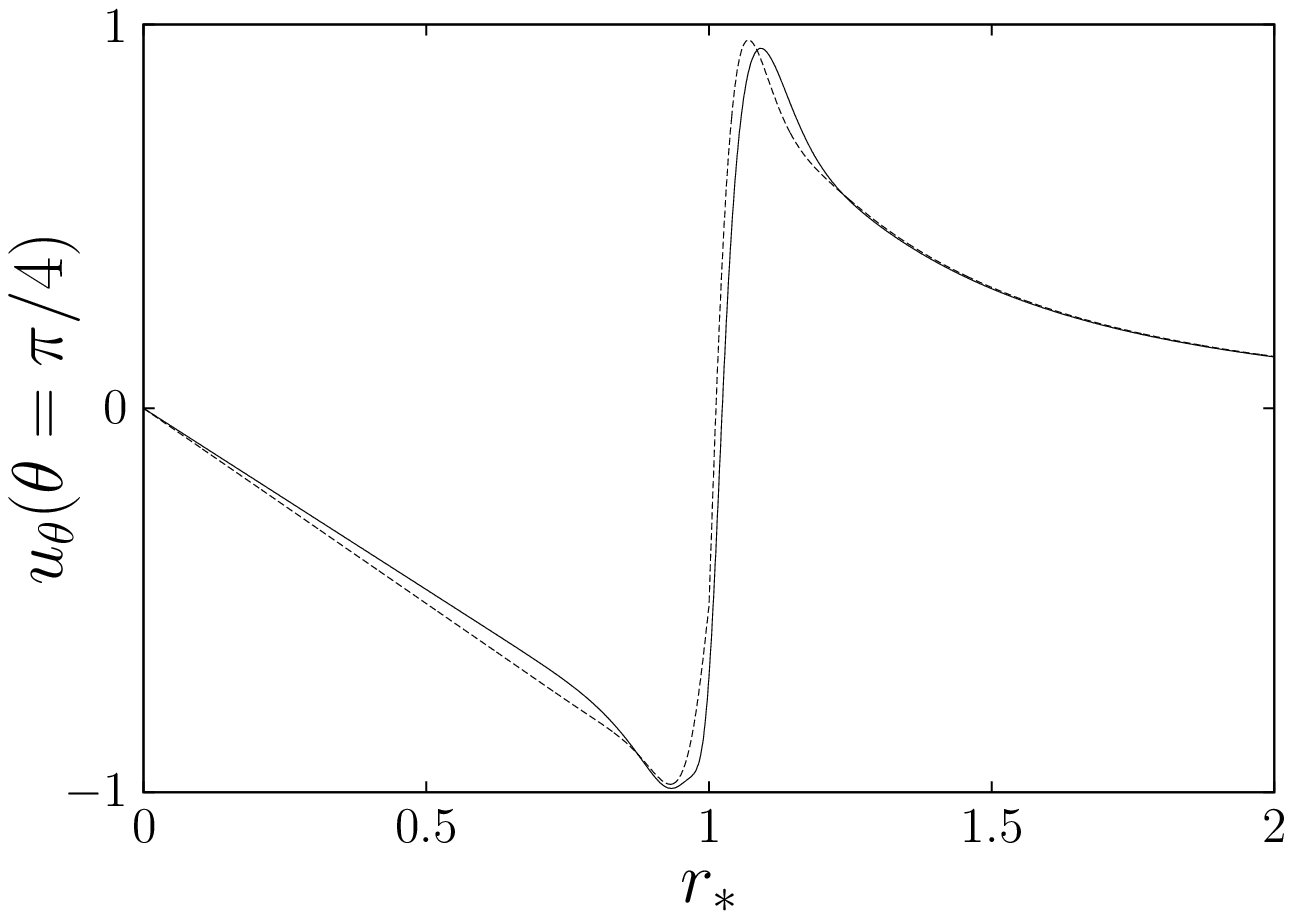}\\[0.25cm]
 \includegraphics[height=4.5cm]{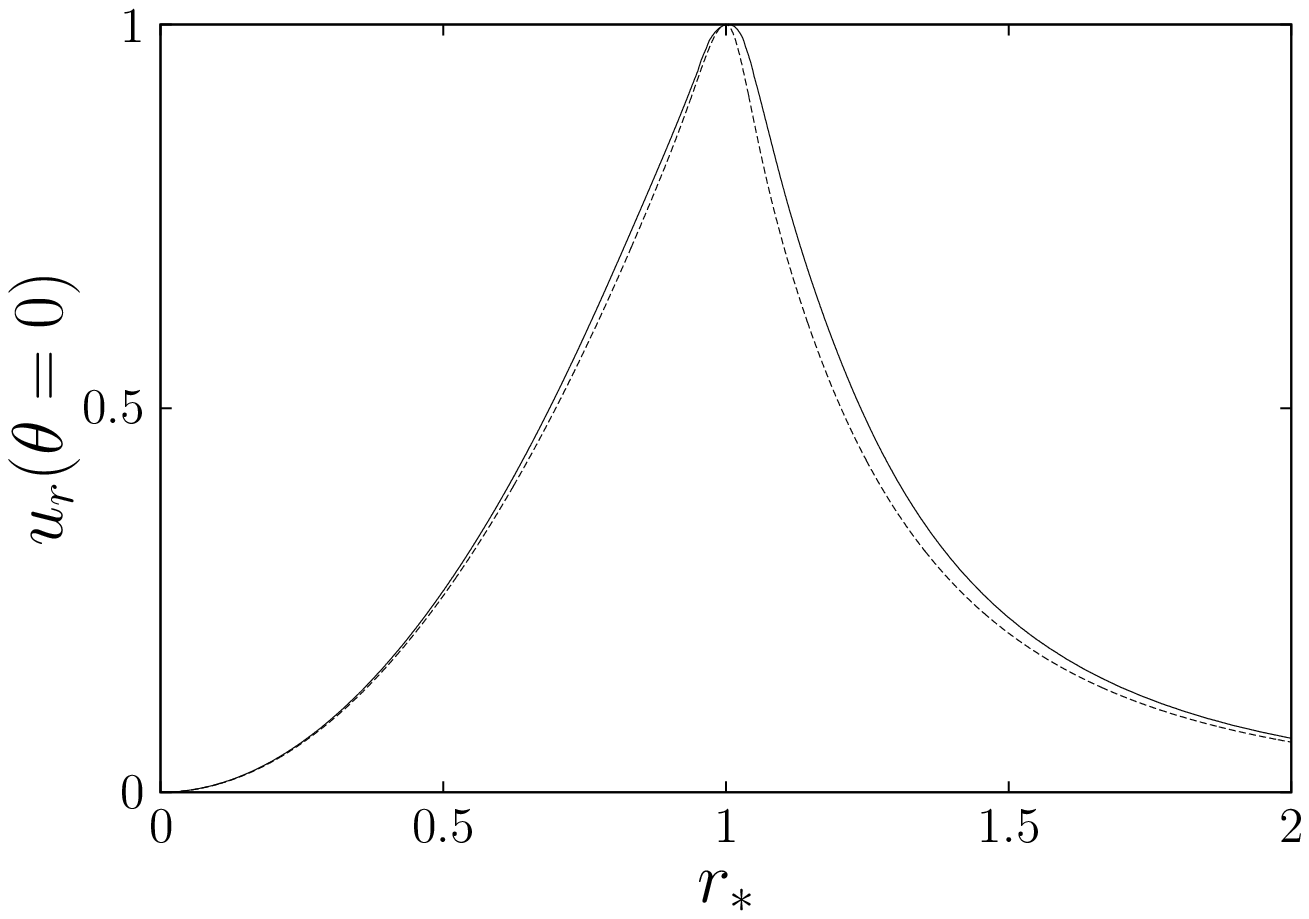}\hfill\includegraphics[height=4.5cm]{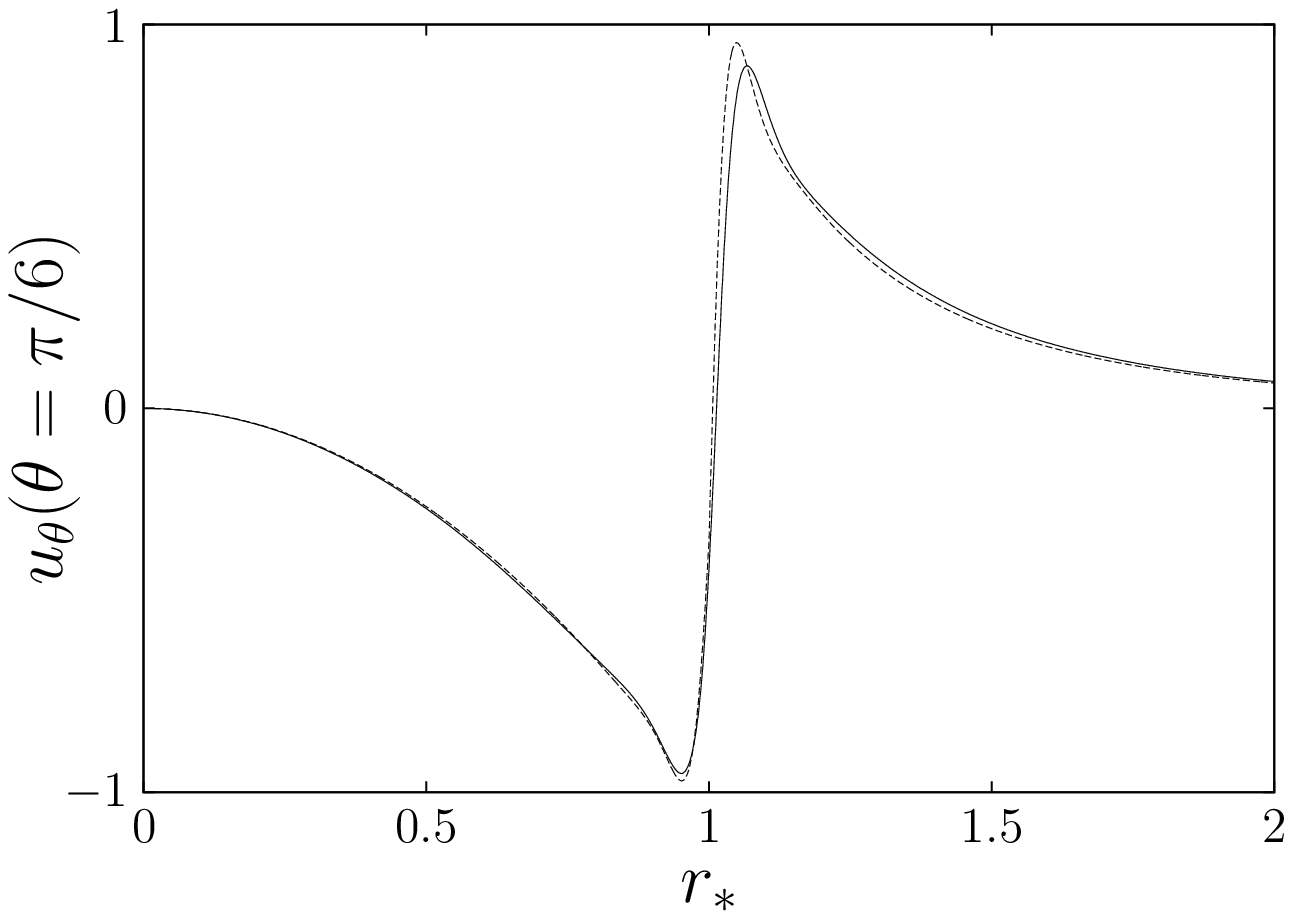}\\[0.25cm]
 \includegraphics[height=4.5cm]{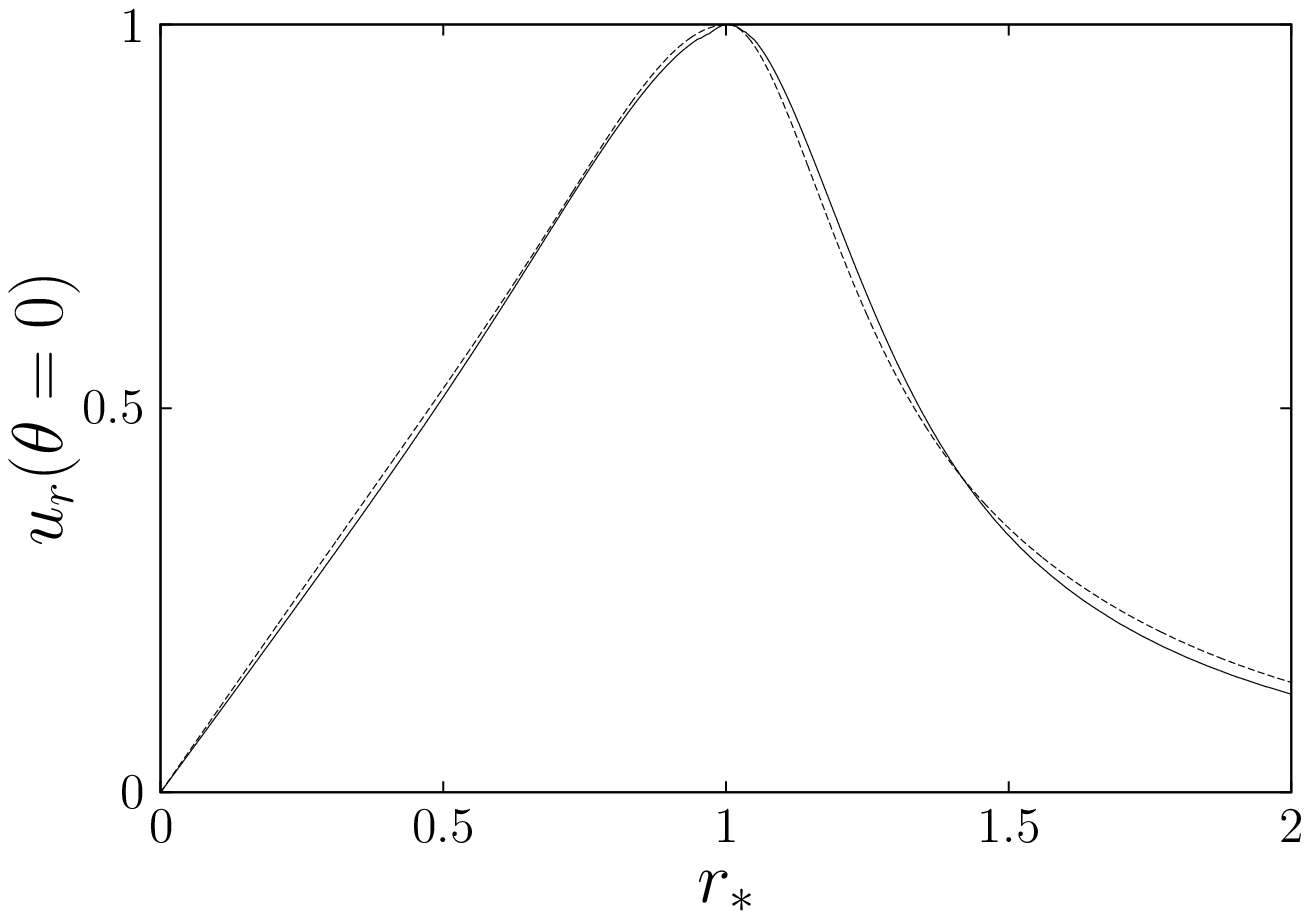}\hfill\includegraphics[height=4.5cm]{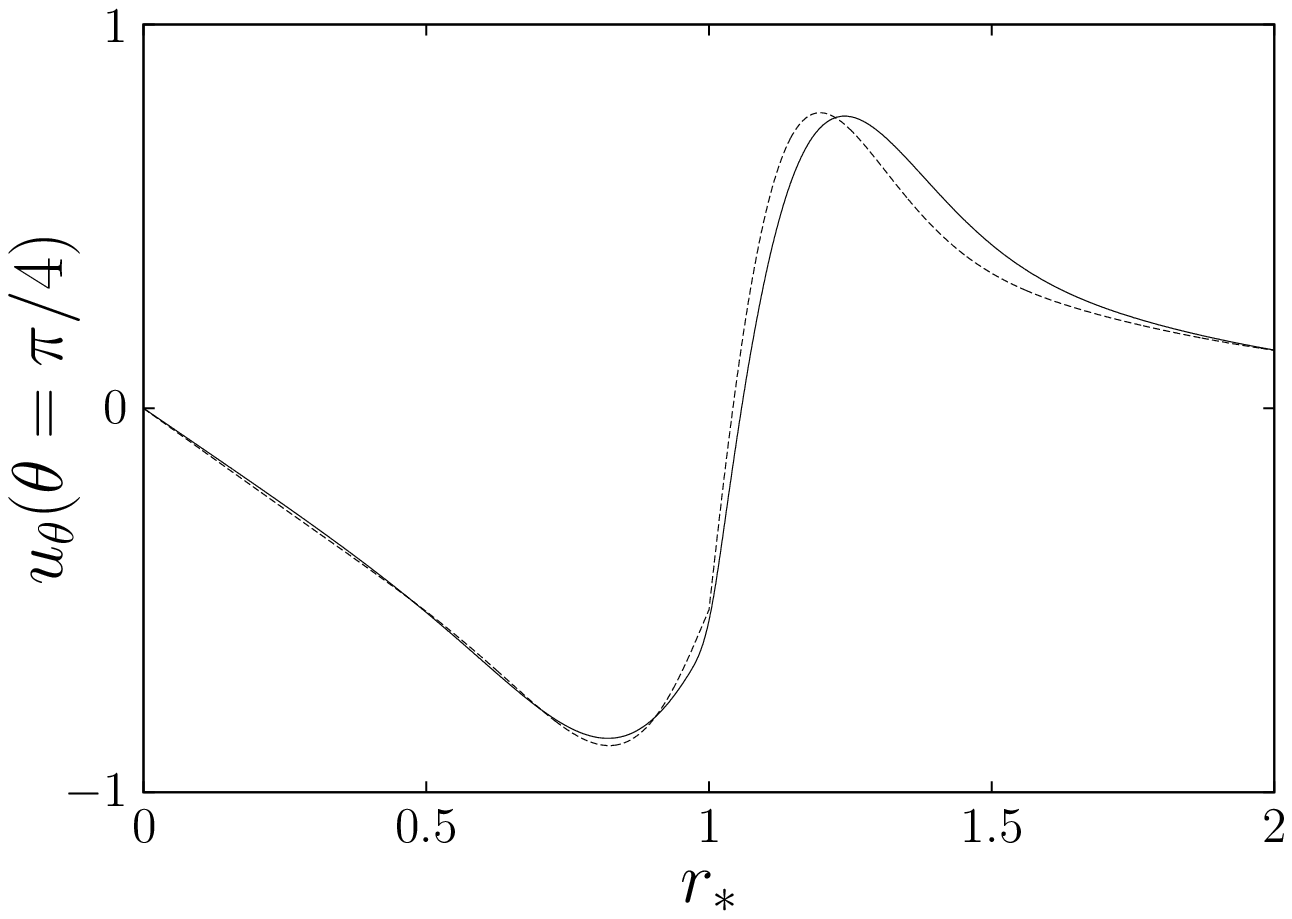}
 \caption{Radial dependence of one component of the fluid velocity,
  captured at one passage time of the membrane through the reference circle,
  from the analytical approach (\dashed) and from numerical simulations (\solid).
  Left panels show the radial velocity at $\theta=0$ (where $u_{\theta}$ is zero within approximation),
  right panels the azimuthal velocity at $\theta=\upi/2n$ (where $u_r$ is zero within approximation).
  Plotted are the ``standard'' case $n=2$ \& $\nu_*=0.001$ \& $R_*=0.5$ (top row),
  the case with different $n=3$ (middle row), and the case with increased $\nu_*=0.01$ (bottom row).}
 \label{compar}
\end{figure}

 After numerically finding $\phi(r_*)$, it is possible to reconstruct the whole
 stream function $\psi(r,\theta,t)$ from (\ref{nom}). Instantaneous sketches of the fluid velocity in an area around
 the membrane are presented in figures \ref{sketch}--\ref{elli} for three situations:
 our reference case, a case with different $n$, and a case with increased $\nu_*$ (see table \ref{tav}).
 The former figure corresponds to the passage of the membrane through the circular state,
 while the latter represents the time instant of maximum membrane eccentricity.
 About this latter state it is interesting to notice that in our formalism, as discussed in the appendix,
 the membrane moves as a block (there exists a countable series of time instants when all its points are contemporaneously at rest),
 but the fluid inertia --- due to the time derivative in (\ref{nsrad}--\ref{nsazi}) ---
 acts so as to maintain some fluid motion almost everywhere at any time, as noticeable in figure \ref{elli}.
 On the contrary, from figure \ref{sketch} (and from the dashed lines of the forthcoming figure \ref{compar}),
 one can conclude that --- when the membrane passes through the circular state --- the maximum velocity gradients
 are located at some distance outside, and that the thickness of the boundary layer grows with $\nu_*$.
\begin{table}
 \begin{center}
  \begin{tabular}{c|ccc|c|c}
   case$\quad$ & $\quad n\ $ & $\ \nu_*\ $ & $\ R_*\quad$ & $\quad\omega_*$ (analytical)$\quad$ & $\quad\omega_*$ (numerical)\\
   \hline
   \#0$\quad$ & $\quad2\ $ & $\ 0.001\ $ & $\ 0.5\quad$ & $\quad1.6951-0.0374\ui\quad$ & $\quad1.6702-0.0396\ui$\\
   \#1$\quad$ & $\quad3\ $ & $\ 0.001\ $ & $\ 0.5\quad$ & $\quad3.3947-0.0713\ui\quad$ & $\quad3.3354-0.0668\ui$\\
   \#2$\quad$ & $\quad2\ $ & $\ 0.01\ $ & $\ 0.5\quad$ & $\quad1.6148-0.1219\ui\quad$ & $\quad1.5772-0.1435\ui$
  \end{tabular}
 \end{center}
 \caption{The set of parametres used for the three situations represented in each row of figures \ref{sketch}--\ref{compar},
  and indicated by different grey filling tones in figure \ref{raddro}. Case \#0 is our standard or reference benchmark.
  Also reported are the values of the nondimensional complex angular frequency, from our linear analysis and from YALES2BIO simulations (see text).}
 \label{tav}
\end{table}

 As a comparison, results from numerical simulations are also provided, obtained by means of the
 in-house YALES2BIO solver developed at Institut de Math\'ematiques et de Mod\'elisation de Montpellier.
 YALES2BIO builds heavily upon the YALES2 solver designed at the CORIA (CNRS UMR 6614), which solves the
 incompressible Navier--Stokes equations using a finite-volume approach \citep{Moureau:2011a,Moureau:2011b}.
 The classical immersed-boundary method, or IBM \citep{Peskin:1977}, is implemented to solve
 the present fluid--structure interaction coupling \citep{Mendez:2013}.
 The capsule is represented as a set of discrete massless markers convected by the flow field,
 and pairwise connected by springs. Given the strain in each spring, the state of membrane stress is obtained,
 and this information is passed to the fluid-flow solver, which solves the Navier--Stokes equations forced by the membrane action.
 As already stated in the introduction, IBM and similar methods have often been used
 to simulate the damped oscillations of an initially-elliptic capsule \citep{Tu:1992,Lee:2003,Song:2008,Tan:2008}.
 Circular capsules are initially deformed with a small perturbation of given azimuthal wave number,
 and plugged in a fluid at rest. The capsule is let evolve and oscillates before reaching the circular equilibrium shape.
 The frequency and damping rate of oscillations are extracted from the temporal evolution of the markers' position.
 First periods are not considered, in order not to spoil the results with effects of the initial conditions.\\
 In figure \ref{compar} we plot the dependence of the velocity field on the radial variable,
 for both the radial and the azimuthal component, in the three situations already mentioned
 and at a passage time through the circle (fixed by means of the procedure reported in the appendix).

\section{Discussion} \label{disc}

 In this section we compare our results to similar cases already studied in literature,
 and we propose a physical model to interpret them.

\subsection{Droplets}

\begin{figure}
 \centering
 \includegraphics[height=6cm]{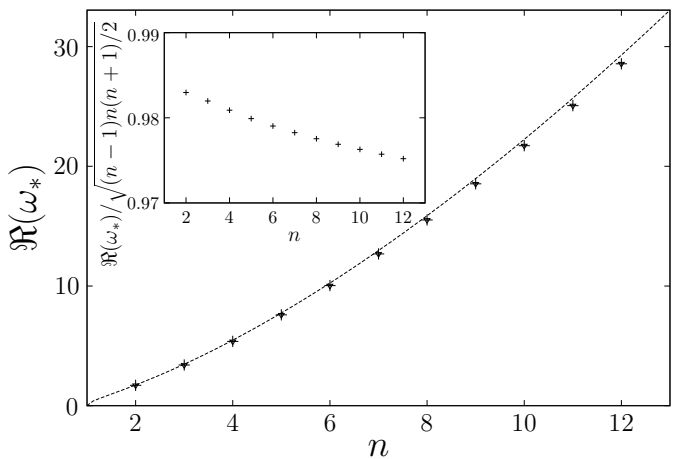}
 \caption{Same as in the left panel of figure \ref{W} but with the imposition of constraint (\ref{non2BIS}) for droplets ($+$).
  For comparison, also the case of membranes with the usual constraint (\ref{non2}) is plotted ($\triangledown$).
  (The difference is almost undistinguishable on this scale, but for the modes $n=2,3$ a small discrepancy can be noticed from the forthcoming figure \ref{raddro}.)
  The dashed line represents the inviscid prediction from (\ref{poi}),
  whose goodness is confirmed by the fact that the ratio plotted in the insert approximately equals 1.}
 \label{mon}
\end{figure}

 A first comparison can be carried out with the case of droplet oscillations.
 \citet{Kelvin:1890} and \citet{Rayleigh:1896} analysed this problem in the presence of a three-dimensional
 inviscid/potential/irrotational flow,
 for which no damping occurs, in the specific cases of cavities or gas bubbles in a liquid or of liquid drops in vacuum.
 \citet{Lamb:1932,Chandrasekhar:1961,Miller:1968} and \citet{Padrino:2008} later removed these limitations successively.
 \citet{Prosperetti:1980a} noticed the presence of a full continuous spectrum of complex frequencies
 --- namely, purely-imaginary ones, corresponding to a pure damping ---
 besides the classical ``isolated'' solution in the complex plane.
 Even if the latter is the only relevant to us, because endowed with non-zero real and imaginary parts,
 through our numerical procedure we also find a hint of the existence of a set of several imaginary solutions,
 which is probably not continuous due to the presence of the external wall
 but which confirms the agreement with the literature even for a different situation.
 \citet{Prosperetti:1980b} also analysed the oscillations of bubbles and drops in terms of the initial-value problem,
 which as already stated is not our aim here.
 Theoretical and experimental results on the role of surfactants were found by \citet{Lu:1991} and \cite{AbiChebel:2012}.\\
 From the aforementioned references, a general expression for the oscillation angular frequency in 3D inviscid flows is
 \begin{equation} \label{agn}
  \Re(\omega)=\left\{\frac{(n-1)n(n+1)(n+2)\zeta}{R^3[n\rho_{\mathrm{ext}}+(n+1)\rho_{\mathrm{int}}]}\right\}^{1/2}\;,
 \end{equation}
 while for 2D one gets
 \begin{equation} \label{poi}
  \Re(\omega)=\left[\frac{(n-1)n(n+1)\zeta}{R^3(\rho_{\mathrm{ext}}+\rho_{\mathrm{int}})}\right]^{1/2}\;,
 \end{equation}
 There, $\zeta$ is the surface tension (equivalent to our $E(R-R_0)/R_0$),
 and the fluids external and internal to the membrane may have any mass densities
 ($\rho_{\mathrm{ext}}$ and $\rho_{\mathrm{int}}$, respectively).
 Therefore, in our notation, (\ref{poi}) becomes:
 \begin{equation} \label{astel}
  \Re(\omega_*)=\sqrt{\frac{(n-1)n(n+1)}{2}}\;.
 \end{equation}
 Cumbersome formulae also exist in literature \citep{Lamb:1932,Miller:1968}
 for the damping rate $\Im(\omega)$ of droplets or bubbles, or for fluids of very high or low viscosities.\\
 In our formalism, a droplet (in another liquid of the same density) can be thought of as a membrane with uniform surface tension,
 which amounts to neglecting the angle-dependent term in (\ref{elref}) (and thus to pose $\chi=\bar{\chi}$ in (\ref{decomzero})).
 As a consequence, equation (\ref{nondi}) and constraint (\ref{non3}) keep the same form,
 but the interfacial jump of the second derivative is now given no longer by (\ref{non2}) but rather by:
 \begin{equation} \label{non2BIS}
  \phi^{(2)-}-\phi^{(2)+}=0\;.
 \end{equation}
 In figure \ref{mon} we show the (real part of) the angular frequency as a function of $n$
 if constraint (\ref{non2BIS}) is imposed, together with the inviscid prediction from (\ref{astel}).
 The approximate constancy ($\simeq1$) of the ratio $\Re(\omega_*)/[(n-1)n(n+1)/2]^{1/2}$,
 depicted in the insert of figure \ref{mon}, proves the excellence of this guess.

\subsection{Radial transport}

\begin{figure}
 \centering
 \includegraphics[height=6cm]{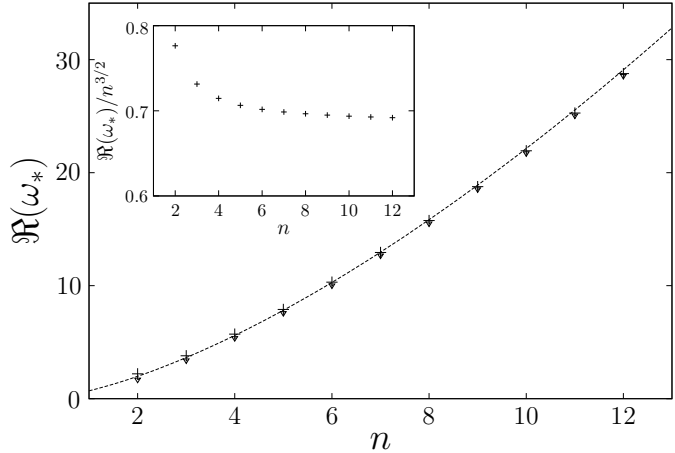}
 \caption{Same as in the left panel of figure \ref{W} but with the imposition of constraint (\ref{non2TER}) for radial transport ($+$).
  For comparison, also the case of complete transport with the usual constraint (\ref{non2}) is plotted ($\triangledown$).
  (The difference is very small on this scale, but for the modes $n=2,3$ a definite discrepancy can be noticed from the forthcoming figure \ref{raddro}.)
  The dashed line represents the prediction from our model (\ref{fit}),
  where the coefficient $\beta=0.7$ has been fitted from the ratio plotted in the insert.}
 \label{cas}
\end{figure}

 It is also interesting to investigate what happens when the membrane can slip
 with respect to the fluid, and the latter transports the former only in the radial direction.
 In this case, the appropriate expression for the elongation is no more the dynamical one (\ref{elref}) but rather a purely-geometric one,
 \[\chi(\theta,t)=\frac{\mathcal{R}(\theta,t)-R_0}{R_0}=\frac{R-R_0}{R_0}+\frac{\Delta(\theta,t)}{R_0}\;.\]
 The consequence of this modification in the forcing terms (\ref{force}) is once again encoded
 in a change of the second-derivative jump, from (\ref{non2}) to
 \begin{equation} \label{non2TER}
  \phi^{(2)-}-\phi^{(2)+}=\frac{\ui n^2}{\omega_*\nu_*(1-R_*)}\phi(1)\;,
 \end{equation}
 while (\ref{nondi}) and (\ref{non3}) maintain the same form. (The consequent results will be described in figure \ref{cas}.)
 We can assert that, with respect to a purely-radial slippery transport, the mechanism of tangential friction and transport
 increases the degree of homogenization of the membrane elongation or tension, towards the limiting case represented by droplets.\\
 We can now propose a physical model to interpret our results, based on the concepts of loaded fluid and vibrating string.
 We can think that the membrane ``embarks'' some fluid during its movement,
 because mass conservation (or, equivalently, the impossibility of creating vacuum pockets)
 forces some fluid to follow the membrane in its radial movement.
 This explains why oscillations can be observed. Indeed, a deformed massless membrane in vacuum
 would asymptotically (i.e., with a vanishing final velocity) reach its equilibrium shape without overpassing it,
 and the same would hold in a fluid described by the pure Stokes equation.
 However, even in the absence of the nonlinear advective contribution from the full Navier--Stokes equation,
 the presence of the time derivative represents a fluid inertial term which justifies the observed multiple overpassing.
 The effective inertia of the membrane can thus be quantified by investigating how much fluid
 is carried per unit membrane surface (here, length), i.e.\ by finding its equivalent thickness.\\
 It is well known that a vibrating string has oscillation angular frequency $\Omega$ and tension
 (per unit length in the third invariant direction $z$, perpendicular to the oscillation plane) $\mathcal{T}$ given by
 \begin{equation} \label{vibra}
  \Omega=\frac{n}{R}\sqrt{\frac{\mathcal{T}}{m}}\;,\qquad\mathcal{T}=E\bar{\chi}=E\frac{R-R_0}{R_0}\;,
 \end{equation}
 where $m$ is the mass per unit string arclength and per unit length in $z$.
 Now, this would usually be given by the mass of a solid material string in vacuum;
 but here the key point is that, due to the massless membrane, this mass is exclusively represented by the fluid,
 so we can express it in terms of an equivalent fluid thickness $d$ as:
 \begin{equation} \label{mass}
  m=\rho d\;.
 \end{equation}
 Inverting (\ref{mass}) and exploiting (\ref{vibra}), we obtain:
 \begin{equation} \label{thick}
  d=\frac{m}{\rho}=\frac{E}{\rho}\frac{R-R_0}{R_0}\left(\frac{n}{R\Omega}\right)^2\;.
 \end{equation}
 Equation (\ref{thick}) allows us to empirically find an effective thickness
 for each case we simulate numerically, by substituting the outcome $\Re(\omega)$ in place of $\Omega$.
 For the cases of purely-radial transport (\ref{non2TER}), our guess is an approximately linear relation
 between $d$ and the wavelength $\lambda=2\upi R/n$. Indeed, if one defines this proportionality factor
 as $\alpha= d/\lambda$, replacing this empirical relation into (\ref{thick}) and nondimensionalising, one gets
 \begin{equation} \label{fit}
  \Re(\omega_*)=\sqrt{\frac{n^3}{2\upi\alpha}}=\beta n^{3/2}\;,
 \end{equation}
 where $\beta=(2\upi\alpha)^{-1/2}$.
 The insert of figure \ref{cas} shows that the ratio $\Re(\omega_*)/n^{3/2}$ from our scheme
 effectively equals the constant value $\beta\simeq0.7$ in a good approximation,
 and such a value is used in figure \ref{cas} to reproduce the fit \ref{fit}.
 As a consequence, $d/\lambda=\alpha=1/2\upi\beta^2\simeq0.3$. (For the general cases of complete transport
 (\ref{non2}), this value ranges from $0.5$ at small $n$ to $0.3$ at large $n$,
 so one may think that in this latter asymptotics the two aforementioned cases tend to confuse with each other.)
 
\subsection{Comparison between different schemes}

\begin{figure}
 \centering
 \includegraphics[height=6cm]{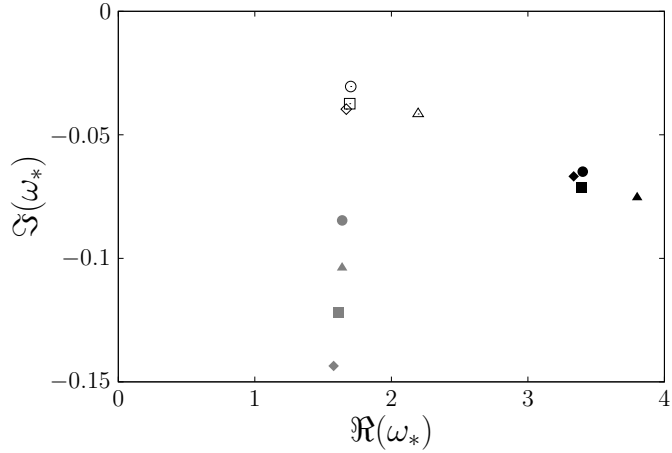}
 \caption{Sketch of the solutions in the complex $\omega_*$ plane for our reference case
  $n=2$ \& $\nu_*=0.001$ \& $R_*=0.5$ (open symbols), together with the case $n=3$ (filled symbols)
  and the case $\nu_*=0.01$ (grey symbols). Squares represent the usual calculations
  with constraint (\ref{non2}), and diamonds their numerical counterparts from YALES2BIO;
  circles represent droplets (constraint (\ref{non2BIS}));
  triangles represent purely-radial transport with slippery membrane (constraint (\ref{non2TER})).}
 \label{raddro}
\end{figure}

 The results for $\omega_*$ corresponding to imposing (\ref{non2}) --- with its numerical counterpart from YALES2BIO ---
 or (\ref{non2BIS}) or (\ref{non2TER}) are plotted in figure \ref{raddro}, for three sets of parametres:
 our reference case, a case with different $n$, and a case with increased $\nu_*$.
 The full CFD approach turns out to be in excellent agreement with the present analytical point of view.
 From the larger spreading of the grey symbols we deduce that, when the reduced viscosity is higher,
 the role of the interface conditions is more relevant, in the sense that the cases of droplets
 and of purely-radial transport become definitely separated from the original capsule one.
 Notice that this trend is not observed when the parametre made larger is the mode index,
 because filled symbols are as spread out as the open ones.

\begin{figure}
 \centering
 \includegraphics[height=6cm]{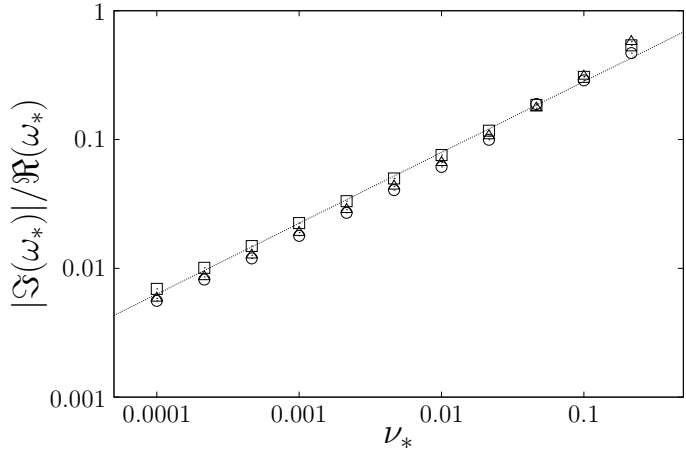}
 \caption{Ratio between the (opposite of the) imaginary part and the real part of the complex frequency
  plotted in figure \ref{ni}. Squares, circles and triangles have the same meaning as in figure \ref{raddro},
  while the dashed line represents the fit $|\Im(\omega_*)|/\Re(\omega_*)=\nu_*^{0.55}$.}
 \label{malo}
\end{figure}

 In particular, it is interesting to investigate the ratio
 between the (absolute value of the) imaginary part and the real one,
 and its behaviour with the reduced viscosity.
 Figure \ref{malo} shows that a power law with exponent 0.55 and unit prefactor
 is an excellent approximation of all three schemes --- membrane, droplet and radial transport.
 This implies that, once an effective inverse time scale is introduced as a reference unit
 via $\Re(\omega)$ and/or expressions (\ref{fit}) \& (\ref{astel}), the damping rate
 in this unit is of course a growing function of $\nu_*$, but is almost independent
 of the implemented physical scheme. In other words, at fixed reduced viscosity,
 the effects of the interfacial boundary layer look negligible to compute $\Im(\omega)$ in the aforementioned rescaled unit.

\section{Conclusions and perspectives} \label{conc}

 We have investigated the damped oscillations of a massless elastic membrane
 immersed in a two-dimensional viscous incompressible fluid.
 By introducing a normal-mode decomposition of the stream function,
 we have analytically obtained a fourth-order linear ordinary differential equation.
 Such an equation possesses non-trivial solutions only for a subset of complex frequencies,
 and namely for a unique value of $\omega$ with non-zero both real (= oscillation angular frequency)
 and imaginary (= damping rate) parts. By means of an appropriate nondimensionalisation
 of the problem, the latter two quantities have been shown to depend on the oscillation mode,
 on the reduced viscosity and on the pre-inflation ratio. A detailed numerical study of these
 behaviours has been performed, as well as an (excellent) comparison with the results of a
 fully-computational procedure we have developed using the YALES2BIO code, in terms of both
 the ``generalised eigenvalues'' $\omega$ and the corresponding eigenvectors
 --- i.e.\ the shape of the generated fluid flow.
 
 An interesting feature of our approach is that it unifies the case of an elastic capsule
 with those of a droplet or of a slippery membrane, in the sense that simply turning on or off
 some parametres in the equations we have been able to reproduce the corresponding results
 from the scientific literature or from the simple physical model we have proposed.
 
 The present work is also the first, to the best knowledge of its authors, to provide
 semi-analytical results in the case of damped oscillations of a two-dimensional capsule
 closed by an elastic membrane, though it has proven to be a popular configuration
 in quality assessment of numerical methods for computing the dynamics of capsules.
 It is anticipated that our results could be used as a quantitative validation for
 future numerical methods.
 
 We thank Etienne Gibaud, Vanessa Lleras, Daniele di Pietro, Matthieu Alfaro and R\'emi Carles for useful discussions and suggestions.
 We acknowledge financial support from LabEx NUMEV (ANR-10-LABX-20), ANR FORCE (ANR-11-JS09-011-01) and Dat\symbol{64}Diag (OSEO).

\appendix

\section{Details of the analytical procedure} \label{appe}

 In this appendix we show in more details the calculations of sections \ref{prob} and \ref{equa}.
 
 In (\ref{force}), the curvilinear derivative following $\hat{\bm{\tau}}$ is:
 \begin{equation} \label{cude}
  \frac{\ud}{\ud\tau}=[(R+\Delta)^2+\Delta'^2]^{-1/2}\frac{\partial}{\partial\theta}\;.
 \end{equation}
 The pairs of polar and curvilinear unit vectors are related to each other by
 \begin{equation} \label{uvtan}
  \hat{\bm{\tau}}=[(R+\Delta)^2+\Delta'^2]^{-1/2}\left[\Delta'\hat{\bm{e}}_r+(R+\Delta)\hat{\bm{e}}_{\theta}\right]
 \end{equation}
 and
 \begin{equation} \label{uvnor}
  \hat{\bm{N}}=[(R+\Delta)^2+\Delta'^2]^{-1/2}\left[-(R+\Delta)\hat{\bm{e}}_r+\Delta'\hat{\bm{e}}_{\theta}\right]\;.
 \end{equation}
 In the limit of small-amplitude deformations, at leading order we have the following simplifications:
 \begin{eqnarray*}
  &&\displaystyle\Gamma(\theta,t)\simeq\frac{1}{R}\left(1-\frac{\Delta}{R}-\frac{\Delta''}{R}\right)\;,\\
  &&\displaystyle\frac{\ud}{\ud\tau}\simeq\frac{1}{R}\left(1-\frac{\Delta}{R}\right)\frac{\ud}{\ud\theta}\;,\\
  &&\displaystyle\hat{\bm{\tau}}\simeq\frac{\Delta'}{R}\hat{\bm{e}}_r+\hat{\bm{e}}_{\theta}\;,\\
  &&\displaystyle\hat{\bm{N}}\simeq-\hat{\bm{e}}_r+\frac{\Delta'}{R}\hat{\bm{e}}_{\theta}\;,\\
  &&\displaystyle\delta(r-\mathcal{R})\simeq\delta(r-R)-\Delta\dot{\delta}(r-R)\;.
 \end{eqnarray*}

 To describe the membrane elongation, let us consider a material element
 of present length $L(t)=||\bm{x}-\bm{x}_{\dag}||=\ud s$, between a present point
 $\bm{x}=(R+\Delta(\theta,t),\theta)$ (in polar coordinates) and its neighbour $\bm{x}_{\dag}=\bm{x}+\hat{\bm{\tau}}\ud s$.
 After an infinitesimal time $\ud t$, these two points are located in $\bm{x}^{\star}=\bm{x}+\bm{u}(\bm{x},t)\ud t$
 and $\bm{x}_{\dag}^{\star}=\bm{x}_{\dag}+[\bm{u}(\bm{x},t)+(\bm{x}_{\dag}-\bm{x})\cdot\nabla\bm{u}(\bm{x},t)]\ud t$,
 respectively, from which one can easily obtain the new length $L(t+\ud t)$.
 Using (\ref{uvtan}), after straightforward algebra, the length increment thus reads
 \begin{equation} \label{incrlen}
  L(t+\ud t)-L(t)=\ud t\,\ud s\left[\frac{\Delta'\displaystyle\frac{\ud u_r}{\ud\tau}+(R+\Delta)\displaystyle\frac{\ud u_{\theta}}{\ud\tau}}{\sqrt{(R+\Delta)^2+\Delta'^2}}+\frac{(R+\Delta)u_r-\Delta'u_{\theta}}{(R+\Delta)^2+\Delta'^2}\right]\;,
 \end{equation}
 where all velocities and their derivatives must be computed in $(R+\Delta(\theta,t),\theta)$.
 Plugging this result into (\ref{elref}), using (\ref{cude}) and eliminating $L_0$, we get:
 \begin{eqnarray} \label{increlon}
  \chi(t+\ud t)-\chi(t)&=&\frac{L(t+\ud t)-L_0}{L_0}-\frac{L(t)-L_0}{L_0}=\frac{L(t+\ud t)-L(t)}{L(t)/[\chi(t)+1]}\\
  &=&\frac{\ud t\,[\chi(t)+1]}{(R+\Delta)^2+\Delta'^2}\left[\Delta'\left(\frac{\partial u_r}{\partial\theta}-u_{\theta}\right)+(R+\Delta)\left(\frac{\partial u_{\theta}}{\partial\theta}+u_r\right)\right]\!.\nonumber
 \end{eqnarray}
 Let us now decompose the elongation into a mean constant part --- averaged on the angle and denoted by a bar ---
 and a fluctuating one (small by definition and denoted by a tilde). We have:
 \begin{equation} \label{decomzero}
  \chi(\theta,t)=\bar{\chi}+\tilde{\chi}(\theta,t)=\frac{R-R_0}{R_0}+(\textrm{term with zero angular average})\;.
 \end{equation}
 Replacing (\ref{decomzero}) into (\ref{increlon}), on the left-hand side only the fluctuating part survives,
 and at leading order on the right-hand side we can neglect both the fluctuating part in $\chi$
 and the quadratically-small terms with $\Delta$ and $\Delta'$ (remember that, in this approach,
 $\bm{u}$ is a small quantity itself). For the same reason, dividing (\ref{increlon}) by $\ud t$,
 we can interpret the material derivative $\ud\tilde{\chi}/\ud t$ as a temporal partial derivative
 $\partial\tilde{\chi}/\partial t$, because of the quadratic smallness of the convective contribution;
 and we can also consider all velocities and their derivatives as computed in $(R,\theta)$. We are left with:
 \begin{equation} \label{gammazero}
  \frac{\partial\tilde{\chi}(\theta,t)}{\partial t}=\frac{\bar{\chi}+1}{R^2}R\left[\frac{\partial u_{\theta}}{\partial\theta}+u_r\right](R,\theta,t)\;.
 \end{equation}

 Passing to the analysis of the equations, by plugging (\ref{stream}) into (\ref{mixed}), we get:
 \begin{eqnarray} \label{psi}
  r^4\frac{\partial^3\psi}{\partial r^2\partial t}+r^2\frac{\partial^3\psi}{\partial\theta^2\partial t}+r^3\frac{\partial^2\psi}{\partial r\partial t}&=&\nu\left(r^4\frac{\partial^4\psi}{\partial r^4}+2r^2\frac{\partial^4\psi}{\partial r^2\partial\theta^2}+\frac{\partial^4\psi}{\partial\theta^4}+2r^3\frac{\partial^3\psi}{\partial r^3}\right.\nonumber\\
  &&\quad\left.-2r\frac{\partial^3\psi}{\partial r\partial\theta^2}-r^2\frac{\partial^2\psi}{\partial r^2}+4\frac{\partial^2\psi}{\partial\theta^2}+r\frac{\partial\psi}{\partial r}\right)\nonumber\\
  &&+\frac{E}{\rho R_0}\left\{\left[\Delta'\left(1-\frac{R_0}{R}\right)\left(1-\frac{r}{R}\right)-\frac{rR_0}{R}\chi'\right]\dot{\delta}(r-R)\right.\nonumber\\
  &&\qquad+\left.\left[\frac{\Delta'''}{R}\left(1-\frac{R_0}{R}\right)-2\frac{R_0}{R}\chi'\right]\delta(r-R)\right\}r^3\;.
 \end{eqnarray}
 Equation (\ref{psi}) is accompanied by an appropriate rewriting of the continuity and boundary conditions (\ref{jump}--\ref{bc}),
 and can be closed thanks to two relations expressing the geometric displacement $\Delta$ and the dynamical elongation $\chi$
 as functions of the fluid quantity $\psi(R)$, as we are going to show.

 The normal-mode decomposition of the membrane quantities, analogously to (\ref{nom}), is:
 \begin{equation} \label{chi}
  \Delta(\theta,t)=\sum_{n=-\infty}^{+\infty}\xi_n\ue^{\ui(n\theta-\omega_nt)}\;,
 \end{equation}
 \begin{equation} \label{decom}
  \tilde{\chi}(\theta,t)=\sum_{n\neq0}\gamma_n\ue^{\ui(n\theta-\omega_nt)}\;.
 \end{equation}
 Here, $n$ is integer and $\omega_n$ complex.
 Both for this angular frequency and for the amplitudes $\xi_n$ and $\gamma_n$,
 we drop the subscript $n$ whenever unambiguous.
 The stream function $\psi$ being defined up to an arbitrary additive constant
 (namely, a space-independent, but possibly time-dependent, term),
 the addend corresponding to $n=0$ shows an undetermined degree of freedom $\phi(r)\mapsto\phi(r)+\textrm{constant}$.
 Due to the area-preservation constraint, this is only related to membrane rotation and angular momentum,
 because it gives rise to an azimuthal angle-independent flow $(u_r=0,u_{\theta}(r,t))$.
 Therefore we decide to neglect the mode $n=0$, as well as the modes $n=\pm1$ which would imply a translation of the center of mass.
 Even if a generic superposition of modes could be dealt with, here we mainly focus on an isolated mode at each time.
 For a single mode, the area preservation cannot hold exactly,
 but it does at working order, i.e.\ neglecting all quadratic amplitudes.\\
 Also notice that, due to the reality condition, the following conjugation relations must hold,
 so as to ensure that $\psi$, $\Delta$ and $\tilde{\chi}$ are real quantities
 (namely, waves travelling in the azimuthal direction) upon summing the pairs of contributions with $n$ and $-n$:
 \[\phi_{-n}=\phi^{\star}_n\;,\qquad\xi_{-n}=\xi^{\star}_n\;,\qquad\gamma_{-n}=\gamma^{\star}_n\;,\qquad\omega_{-n}=-\omega^{\star}_n\;.\]
 In fact, in this specific case, the symmetry is even stronger due to the parity of (\ref{psi})
 in $\theta$ (and consequently of the upcoming equation (\ref{phi}) in $n$),
 so that already \emph{for each single} $n$ any solution $\omega_n$
 has a twin solution with same imaginary part and opposite real one.
 As a consequence, the sum of the contributions from $n$ and $-n$ in the series (\ref{nom})\&(\ref{chi}--\ref{decom})
 actually consists of four addends, i.e.\ of two azimuthal waves travelling in opposite directions
 with the same speed and damping rate, which therefore interfere to produce a \emph{stationary wave}.
 For instance,
 \begin{eqnarray} \label{deltasw}
  \Delta(\theta,t)&=&\sum_{n\in\mathbb{N}}\left\{\xi_n\left[\ue^{\ui(n\theta-\omega_nt)}+\ue^{\ui(-n\theta-\omega_nt)}\right]+\xi^{\star}_n\left[\ue^{\ui(-n\theta+\omega^{\star}_nt)}+\ue^{\ui(n\theta+\omega^{\star}_nt)}\right]\right\}\nonumber\\
  &=&4\sum_{n\in\mathbb{N}}\cos(n\theta)\left[\Re(\xi_n)\cos(\Re(\omega_n)t)+\Im(\xi_n)\sin(\Re(\omega_n)t)\right]\ue^{\Im(\omega_n)t}\nonumber\\
  &=&4\sum_{n\in\mathbb{N}}\sqrt{\Re(\xi_n)^2+\Im(\xi_n)^2}\cos(n\theta)\cos(\Re(\omega_n)t+\sigma_{\xi})\ue^{\Im(\omega_n)t}\;,
 \end{eqnarray}
 so that e.g.\ the mode $n=2$ reproduces the fact that the absolute value
 of the displacement is maximum on the axes (antinodes at $\theta=0,\upi/2,\ldots$)
 and zero on the bisectors (nodes at $\theta=\upi/4,3\upi/4,\ldots$).
 The presence of the phase shift $\sigma_{\xi}=\arctan[-\Im(\xi_n)/\Re(\xi_n)]$
 is related to the lack of the initial-value problem and to the MATLAB procedure for linear algebra.
 Figures \ref{sketch} and \ref{compar} consist in an instantaneous capture
 of the fluid flow at time $t=(3\upi/2-\sigma_{\xi})/\Re(\omega_n)$,
 where the phase shift can easily be computed numerically.
 Viz., $\sigma_{\xi}=-0.0184$ for the standard case $n=2$ \& $\nu_*=0.001$ \& $R_*=0.5$,
 $\sigma_{\xi}=-0.0209$ for the case with $n=3$, and $\sigma_{\xi}=-0.0578$ for the case with $\nu_*=0.01$.

 Substituting (\ref{nom}) and (\ref{chi}--\ref{decom}) into (\ref{psi}),
 we are left with an ordinary differential equation for $\phi$ for each value of $n$,
 namely a fourth-order linear homogeneous one:
 \begin{eqnarray} \label{phi}
  \phi^{(4)}+\frac{2}{r}\phi^{(3)}+\left(-\frac{2n^2+1}{r^2}+\frac{\ui\omega}{\nu}\right)\phi^{(2)}&&\nonumber\\
  +\left(\frac{2n^2+1}{r^3}+\frac{\ui\omega}{\nu r}\right)\phi^{(1)}+\left(\frac{n^4-4n^2}{r^4}-\frac{\ui\omega n^2}{\nu r^2}\right)\phi&&\nonumber\\
  +\frac{\ui nE}{\mu R_0}
  \left\{\left[\left(1-\frac{R_0}{R}\right)\left(1-\frac{r}{R}\right)\xi-\frac{rR_0}{R}\gamma\right]\dot{\delta}(r-R)\right.&&\nonumber\\
  -\left.\frac{1}{R}\left[n^2\left(1-\frac{R_0}{R}\right)\xi+2R_0\gamma\right]\delta(r-R)\right\}&=&0\;.
 \end{eqnarray}
 This must be supplemented with an appropriate rewriting of the constraints (\ref{jump}) and (\ref{bc})
 in terms of the stream function and then of $\phi$. The boundary conditions at the origin and at the (circular) wall now read
 \begin{eqnarray*}
   \left.\frac{\phi}{r}\ui n\ue^{\ui(n\theta-\omega t)}\right|_{r=0}&=0=&\left.\left(\frac{\phi^{(1)}}{r}-\frac{\phi}{r^2}\right)\ui n\ue^{\ui(n\theta-\omega t)}\right|_{r=0}\;,\\
   \left.\frac{\phi}{r}\ui n\ue^{\ui(n\theta-\omega t)}\right|_{r=W}&=0=&\left.-\phi^{(1)}\ue^{\ui(n\theta-\omega t)}\right|_{r=W}\;,
 \end{eqnarray*}
 and give rise to (\ref{bcf}).

 Then we have to remember that the membrane moves with the same local and instantaneous value of the fluid velocity.
 In our Eulerian formalism, the full Lagrangian motion must be projected on the radial direction, i.e.:
 \begin{eqnarray} \label{xi}
  &&\frac{\partial\mathcal{R}(\theta,t)}{\partial t}=u_r(R,\theta,t)\ \Longrightarrow\ \frac{\partial\Delta(\theta,t)}{\partial t}=\frac{1}{R}\frac{\partial\psi(R,\theta,t)}{\partial\theta}\nonumber\\
  &&\Longrightarrow\ -\ui\sum_{n=-\infty}^{+\infty}\omega_n\xi_n\ue^{\ui(n\theta-\omega_nt)}=\frac{\ui}{R}\sum_{n=-\infty}^{+\infty}n\phi_n(R)\ue^{\ui(n\theta-\omega_nt)}\nonumber\\
  &&\Longrightarrow\ \xi_n=-\frac{n}{R\omega_n}\phi_n(R)\;.
 \end{eqnarray}
 Analogously, from (\ref{gammazero}), for the elongation we have:
 \begin{eqnarray} \label{gamma}
  &&\frac{\partial\tilde{\chi}(\theta,t)}{\partial t}=\frac{1}{R_0R}\left.\left(\frac{\partial}{\partial\theta}-R\frac{\partial^2}{\partial r\partial\theta}\right)\psi\right|_{(R,\theta,t)}\nonumber\\
  &&\Longrightarrow\ -\ui\sum_{n}\omega_n\gamma_n\ue^{\ui(n\theta-\omega_nt)}=\frac{\ui}{R_0R}\sum_{n}n\left.[\phi_n-R\phi^{(1)}_n]\right|_R\ue^{\ui(n\theta-\omega_nt)}\nonumber\\
  &&\Longrightarrow\ \gamma_n=-\frac{n}{R_0R\omega_n}[\phi_n(R)-R\phi^{(1)}_n(R)]\;.
 \end{eqnarray}
 Similarly to (\ref{deltasw}) and consistently with (\ref{xi}), the stream function modes interfere into a stationary form:
 \begin{eqnarray} \label{psisw}
  \psi(r,\theta,t)&=&\sum_{n\in\mathbb{N}}\left\{\phi_n(r)\left[-\ue^{\ui(n\theta-\omega_nt)}+\ue^{\ui(-n\theta-\omega_nt)}\right]+\phi^{\star}_n(r)\left[-\ue^{\ui(-n\theta+\omega^{\star}_nt)}+\ue^{\ui(n\theta+\omega^{\star}_nt)}\right]\right\}\nonumber\\
  &=&-4\sum_{n\in\mathbb{N}}\sin(n\theta)\left[\Re(\phi_n(r))\sin(\Re(\omega_n)t)-\Im(\phi_n(r))\cos(\Re(\omega_n)t)\right]\ue^{\Im(\omega_n)t}\nonumber\\
  &=&-4\sum_{n\in\mathbb{N}}\sqrt{\Re(\phi_n(r))^2+\Im(\phi_n(r))^2}\sin(n\theta)\sin(\Re(\omega_n)t+\sigma_{\phi}(r))\ue^{\Im(\omega_n)t}\;.
 \end{eqnarray}
 The phase shift $\sigma_{\phi}(r)=\arctan[-\Im(\phi_n(r))/\Re(\phi_n(r))]$
 can easily be computed numerically at $r=R$:
 $\sigma_{\phi}(R)=0.00495$ for the standard case $n=2$ \& $\nu_*=0.001$ \& $R_*=0.5$,
 $\sigma_{\phi}(R)=0.00249$ for the case with $n=3$, and $\sigma_{\phi}(R)=0.0179$ for the case with $\nu_*=0.01$.
 Figure \ref{elli} consists in an instantaneous capture
 of the fluid flow at time $t=(2\upi-\sigma_{\phi}(R))/\Re(\omega_n)$,
 when the membrane has its maximum eccentricity along the $x$ axis and is at rest.
 Of course one would find the same result by time-deriving (\ref{deltasw})
 and by defining $\sigma_{\omega}=\arctan[-\Im(\omega_n)/\Re(\omega_n)]$,
 with $\sigma_{\phi}(R)=\sigma_{\xi}+\sigma_{\omega}$.

 Replacing (\ref{xi}--\ref{gamma}) into (\ref{phi}), we get the equation
 \begin{eqnarray} \label{closed}
  \phi^{(4)}(r)+\frac{2}{r}\phi^{(3)}(r)+\left(-\frac{2n^2+1}{r^2}+\frac{\ui\omega}{\nu}\right)\phi^{(2)}(r)&\nonumber\\
  +\left(\frac{2n^2+1}{r^3}+\frac{\ui\omega}{\nu r}\right)\phi^{(1)}(r)+\left(\frac{n^4-4n^2}{r^4}-\frac{\ui\omega n^2}{\nu r^2}\right)\phi(r)&\nonumber\\
  +\frac{\ui n^2E}{\mu R_0R\omega}\left\{\left[\frac{1}{R}\left[n^2\left(1-\frac{R_0}{R}\right)+2\right]\phi(R)-2\phi^{(1)}(R)\right]\frac{\delta(r-R)}{r}\right.&\nonumber\\
  -\left[\left(1-\frac{R_0}{R}-2\frac{r}{R}+\frac{rR_0}{R^2}\right)\phi(R)+r\phi^{(1)}(R)\right]\frac{\dot{\delta}(r-R)}{r}\bigg\}&=&0
 \end{eqnarray}
 (closed in $\phi$), which can be either translated into the system (\ref{each}) \& (\ref{mem2}--\ref{mem3}),
 or nondimensionalised into:
 \begin{eqnarray} \label{nond}
  \phi^{(4)}(r_*)+\frac{2}{r_*}\phi^{(3)}(r_*)+\left(-\frac{2n^2+1}{r_*^2}+\frac{\ui\omega_*}{\nu_*}\right)\phi^{(2)}(r_*)&&\nonumber\\
  +\left(\frac{2n^2+1}{r_*^3}+\frac{\ui\omega_*}{\nu_*r_*}\right)\phi^{(1)}(r_*)+\left(\frac{n^4-4n^2}{r_*^4}-\frac{\ui\omega_*n^2}{\nu_*r_*^2}\right)\phi(r_*)&&\nonumber\\
  +\frac{\ui n^2}{\omega_*\nu_*(1-R_*)r_*}\left\{\left[\left[n^2\left(1-R_*\right)+2\right]\phi(1)-2\phi^{(1)}(1)\right]\delta(r_*-1)\right.&&\nonumber\\
  -\left.\left[\left(1-R_*-2r_*+r_*R_*\right)\phi(1)+r_*\phi^{(1)}(1)\right]\dot{\delta}(r_*-1)\right\}&=&0\;.
 \end{eqnarray}
 Equation (\ref{nond}) then gives rise to (\ref{nondi}) in both fluid domains,
 and to constraints (\ref{non2}--\ref{non3}). These latter are obtained via a single or double integration from $r_*=1_-$ to $1_+$.
 Notice in particular the relation
 \[\lim_{\epsilon\to0}\int_{1-\epsilon}^{1+\epsilon}\ud r_{\dag}\int_{1-\epsilon}^{r_{\dag}}\ud r_*\,\frac{1}{r_*}\dot{\delta}(r_*-1)=1\;.\]

 All our nomenclature and notation is recalled in tables \ref{tabl}--\ref{toa}.
\begin{table}
 \begin{center}
  \begin{tabular}{rcl}
   letter & definition & description\\[0.1cm]
   $d$ && equivalent fluid thickness for vibrating string\\
   $E$ && membrane thickness times Young modulus (constant)\\
   $\hat{e}_r$ && radial unit vector\\
   $\hat{e}_{\theta}$ && azimuthal unit vector\\
   $\bm{f}(r,\theta,t)$ && line force per unit volume exerted by membrane on fluid\\
   $f_r(r,\theta,t)$ && radial component of $\bm{f}$\\
   $f_{\theta}(r,\theta,t)$ && azimuthal component of $\bm{f}$\\
   $g(n,\nu_*,R_*)$ & $\Im(\omega_*)$ & damping rate of membrane oscillation\\
   $h(n,\nu_*,R_*)$ & $\Re(\omega_*)$ & angular frequency of membrane oscillation\\
   $\Im$ && imaginary part\\
   $L(\theta,t)$ && local instantaneous length of infinitesimal membrane arc\\
   $L_0$ && constant value of $L$ in deflated state\\
   $M$ && number of discretisation points in radial direction\\
   $m$ && mass (per unit arclength and $z$) of vibrating string\\
   $\hat{\bm{N}}$ && unit vector inward normal to membrane\\
   $n$ && mode index\\
   $p(r,\theta,t)$ && fluid pressure\\
   $R$ && radius of reference circle (inflated membrane at rest)\\
   $R_0$ && radius of deflated circular membrane\\
   $R_*$ & $R_0/R$ & deflation degree (inverse of inflation ratio)\\
   $\bm{r}$ && space variable\\
   $r$ && radial coordinate\\
   $r_*$ & $r/R$ & nondimensionalised radial coordinate\\
   $\mathcal{R}(\theta,t)$ && local instantaneous membrane location\\
   $\Re$ && real part\\
   $s$ && curvilinear abscissa\\
   $T$ & $\sqrt{\rho R^3R_0/E(R-R_0)}$ & membrane characteristic time\\
   $t$ && time variable\\
   $\mathcal{T}$ && tension (per unit length in $z$) of vibrating string\\
   $\bm{u}(r,\theta,t)$ && fluid velocity\\
   $u_r(r,\theta,t)$ && radial component of flow field\\
   $u_{\theta}(r,\theta,t)$ && azimuthal component of flow field\\
   $W$ && radius of circular external wall\\
   $x$ && first Cartesian coordinate in oscillation plane\\
   $y$ && second Cartesian coordinate in oscillation plane\\
   $z$ && third invariant direction (orthogonal to oscillation plane)\\
  \end{tabular}
 \end{center}
 \caption{Latin nomenclature.}
 \label{tabl}
\end{table}
\begin{table}
 \begin{center}
  \begin{tabular}{rcl}
   symbol & definition & description\\[0.1cm]
   $\alpha$ & $d/\lambda$ &\\
   $\beta$ & $1/\sqrt{2\upi\alpha}$ &\\
   $\Gamma(\theta,t)$ && local instantaneous membrane curvature\\
   $\gamma_n$ or $\gamma$ && amplitude of $\tilde{\chi}$ in mode $n$\\
   $\Delta(\theta,t)$ & $\mathcal{R}(\theta,t)-R$ & local instantaneous membrane departure from circle\\
   $\theta$ && azimuthal angle\\
   $\lambda$ & $2\upi R/n$ & wavelength\\
   $\mu$ & $\rho\nu$ & constant fluid dynamic viscosity\\
   $\nu$ && constant fluid kinematic viscosity\\
   $\nu_*$ & $\nu T/R^2$ & reduced viscosity\\
   $\rho$ && constant fluid mass density\\
   $\sigma_{\phi}(r)$ & $\arctan[-\Im(\phi)/\Re(\phi)]|_r$ & phase shift of stream function amplitude\\
   $\sigma_{\xi}$ & $\arctan[-\Im(\xi)/\Re(\xi)]$ & phase shift of membrane excess radius amplitude\\
   $\sigma_{\omega}$ & $\arctan[-\Im(\omega)/\Re(\omega)]$ & phase shift of complex angular frequency\\
   $\hat{\bm{\tau}}$ && unit vector counterclockwise tangential to membrane\\
   $\phi_n(r)$ or $\phi(r)$ && amplitude of $\psi$ in mode $n$\\
   $\chi(\theta,t)$ && local instantaneous membrane elongation\\
   $\bar{\chi}$ && constant mean membrane elongation (averaged on $\theta$)\\
   $\tilde{\chi}(\theta,t)$ && local instantaneous fluctuation of membrane elongation\\
   $\psi(r,\theta,t)$ && fluid stream function\\
   $\xi_n$ or $\xi$ && amplitude of $\Delta$ in mode $n$\\
   $\Omega$ && oscillation angular frequency of vibrating string\\
   $\omega_n$ or $\omega$ && complex angular frequency in mode $n$\\
   $\omega_*$ & $\omega T$ & nondimensionalised complex angular frequency\\
   $_{\mathrm{ext}}$ && relative to external fluid\\
   $_{\mathrm{int}}$ && relative to internal fluid\\
   $'$ & $\partial/\partial\theta$ & angular derivative\\
   $\dot{}$ && derivative with respect to argument\\
   $^{(l)}$ & $\partial^l/\partial r^l$ & $l$-th order radial derivative\\
   $^+$ & $\lim_{r\to R_+}$ & limit of fluid quantity at interface from exterior\\
   $^-$ & $\lim_{r\to R_-}$ & limit of fluid quantity at interface from interior\\
   $^{\star}$ && complex conjugation
  \end{tabular}
 \end{center}
 \caption{Greek nomenclature and notation.}
 \label{toa}
\end{table}

 \bibliographystyle{jfm}
 \bibliography{./references0}
\end{document}